\documentclass[a4paper, 10pt]{article}

\usepackage{latexsym}
\usepackage{stmaryrd}

\usepackage{vmargin}
\setpapersize{A4}
\setmarginsrb{35mm}{20mm}{25mm}{15mm}{12pt}{11mm}{0pt}{11mm}

\usepackage{amssymb}

\usepackage[dvips]{graphicx}

\usepackage[scaled=.90]{helvet}
\usepackage{courier}
\usepackage[T1]{fontenc}
\normalfont
\usepackage{mathptmx}

\newcommand{\PAR}{\bindnasrepma}
\newcommand{\TENS}{\otimes}

\def\mathdef{\mathop{\rm def} \nolimits}

\def\src{\mathop{\rm src} \nolimits}
\def\tgt{\mathop{\rm tgt} \nolimits}

\def\stripPT{\mathop{\rm strp_{\TENS \PAR}} \nolimits}

\def\fml{\mathop{\rm fml} \nolimits}
\def\lnk{\mathop{\rm lnk} \nolimits}
\def\tpex{\mathop{{\rm ex}_{\TENS \PAR}} \nolimits}

\def\MLLFml{\mathop{\rm MLLFml} \nolimits}

\def\id{\mathop{\rm id} \nolimits}
\def\ID{\mathop{\rm ID} \nolimits}
\def\con{\mathop{\rm con} \nolimits}

\newtheorem{lemma}{Lemma}
\newtheorem{theorem}{Theorem}
\newtheorem{definition}{Definition}
\newtheorem{corollary}{Corollary}
\newtheorem{proposition}{Proposition}
\newtheorem{example}{Example}
\newtheorem{claim}{Claim}
\newtheorem{subclaim}{Subclaim}

\newenvironment{remark}{\begin{flushleft}{\it Remark.} \ \ }{\end{flushleft}}
\newenvironment{proof}{\begin{flushleft}{\it Proof.} \ \ }{\end{flushleft}}
\newenvironment{ack}{\begin{flushleft}{\bf Acknowledgements.} \ \ }{\end{flushleft}}

\begin{document} 
\title{A Coding Theoretic Study on MLL Proof Nets}
\author{Satoshi Matsuoka\\
National Institute of Advanced Industrial Science and Technology (AIST),\\
1-1-1 Umezono,\\
Tsukuba, Ibaraki,\\
305-8563 Japan \\
{\tt matsuoka@ni.aist.go.jp}}
\maketitle
\begin{abstract}
Coding theory is useful for real world applications.
A notable example is digital television. 
Basically, coding theory is to study a way of detecting and/or 
correcting data that may be true or false. 
Moreover coding theory is an area of mathematics, in which 
there is an interplay between many branches of mathematics, e.g., abstract algebra, combinatorics, 
discrete geometry, information theory, etc.
In this paper we propose a novel approach for 
analyzing proof nets of Multiplicative Linear Logic (MLL)
by coding theory. We define families of 
proof structures and introduce a metric space for
each family. In each family, 
\begin{enumerate}
\item [1.] an MLL proof net is a true code element, and 
\item [2.] a proof structure that is not an MLL proof net is
a false (or corrupted) code element.
\end{enumerate}
The definition of our metrics reflects the duality of the multiplicative connectives elegantly. 
In this paper we show that in the framework one error-detecting is
possible but one error-correcting not. 
Our proof of the impossibility of one error-correcting is interesting 
in the sense that 
a proof theoretic property is proved using a graph theoretic argument. 
In addition, we show that affine logic and MLL + MIX are not appropriate
for this framework. That explains why MLL is better than such similar
logics.
\end{abstract}

\section{Introduction} \label{Introduction}
\begin{figure}[htbp]
\begin{center}
\includegraphics[scale=.5]{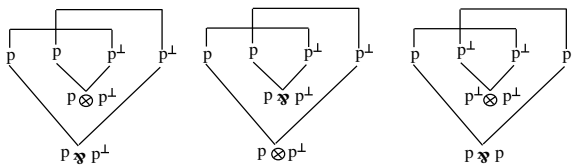}
\caption[An explanatory example]{An explanatory example}  
\label{explanatoryEx}
\end{center}
\end{figure}

The study of the multiplicative fragment of Linear Logic without 
multiplicative constants (for short MLL) \cite{Gir87} is successful from both
semantical and syntactical point of view. 
In semantical point of view  there are good semantical models including 
coherent spaces. In syntactical point of view 
the theory of MLL proof nets has obtained a firm status without doubt.
On the other hand the intuitionistic multiplicative fragment of
Linear Logic without multiplicative constants (for short IMLL) is also
studied,
for example, in \cite{Mat07}. 
IMLL can be seen as a subsystem of MLL. 
IMLL is easier to be studied more deeply than MLL, because we can use
intuitions inspired from the conventional lambda-calculus theory as well as 
graph-theoretic intuitions from the MLL proof nets theory.
We exploited both benefits in \cite{Mat07}. \\
In order to study MLL more deeply, how should we do?
One approach is to interpret MLL intuitionistically by
using G\"{o}del's double negation interpretation. 
One example is \cite{Has05}. However in such an approach 
multiplicative constants must be introduced.  
Definitely introducing multiplicative constants makes things complicated.
Another approach we propose in this paper is to adopt {\it coding theoretic}
framework. \\
Basically, coding theory \cite{Bay98,MS93} is to study a way of detecting and/or 
correcting data that may be true or false. 
Moreover coding theory is an area of mathematics, in which 
there is an interplay between many branches of mathematics, e.g., abstract algebra, combinatorics, 
discrete geometry, information theory, etc.
In this paper we propose a novel approach for 
analyzing proof nets of Multiplicative Linear Logic (MLL)
by coding theory. 
We define families of 
proof structures and introduce a metric space for
each family. In each family, 
\begin{enumerate}
\item [1.] an MLL proof net is a true code element, which is usually called a {\it codeword} in the literature of coding theory;
\item [2.] a proof structure that is not an MLL proof net is
a false (or corrupted) code element.
\end{enumerate}
Figure~\ref{explanatoryEx} shows an explanatory example. 
All three examples in Figure~\ref{explanatoryEx} are MLL proof nets in a standard notation of \cite{Gir87}. 
In our framework the left and the middle proof nets belong to the same family, because 
when we forget $\TENS$ and $\PAR$ symbols, these are the same 
(although in fact, these are equal without forgetting
those symbols. We will discuss the matter later). 
But the right proof net does not belong to the family, because 
when we forget $\TENS$ and $\PAR$ symbols from the right one, we can not identify this one with 
the previous one by the mismatch of the literals $p$ and $p^\bot$. 
The subtle point will be discussed later in a more precise way (see Subsection~\ref{secOurFramework}). 
The definition of our metrics reflects the duality of the multiplicative connectives elegantly. 
Moreover introducing the framework makes it possible to apply different results and techniques of 
other branches of mathematics 
to the study of MLL proof nets. 
In particular, our concern is closely related to the following question: 
given a condition about proof nets (for example, that of the number of ID-links), 
how many proof nets do we have such that they satisfy the condition? 
As far as we know, in the literature, there are only a few discussions about such a counting problem on proof theory. 
\\
So far, most of the study of MLL proof nets have focused on individual proof nets 
(e.g., sequentialization theorem \cite{Gir87}) or 
the relationship between identifiable proof nets (e.g., cut-elimination and $\eta$-expansion). 
On the other hand, our approach focuses on a relationship between similar, but different proof nets. 
In particular, our notion of similarity of proof nets seems to be unable to be understood by conventional type theory.  
\\
The main technical achievement of this paper is Theorem~\ref{mainTheorem2}, which says that 
in our framework one error-detecting is
possible but one error-correcting not. 
Our proof of the theorem is interesting in the sense that 
a proof-theoretic property is proved by a graph-theoretic argument.

\paragraph*{The Structure of the Paper:} Section~\ref{secMLL} introduces basic properties of MLL proof nets. 
MLL proof nets are defined and 
sequentialization theorem on them is described. 
Moreover, the notion of empires, which are needed in order to prove the main theorems, is introduced.  
Section~\ref{secPSFamilies} introduces the notion of PS-families (families of proof structures) 
and distances on them. It is shown that they are metric spaces. 
Then  other basic properties w.r.t PS-families and the main theorems are stated. 
Most of details of the proofs of the main theorems are put into Appendices. 
An example is also given (Example~\ref{exMainEx1}). 
Finally, future research directions about PS-families and elementary results on them are stated.

\section{The MLL System} \label{secMLL}
\subsection{The Basic Theory of MLL Proof Nets}
In this section, we present multiplicative proof nets.
We also call these {\it MLL proof nets} (or simply, {\it proof nets}).
First we define MLL formulas.  In this paper, we only consider MLL formulas 
with the only one propositional variable $p$ because the restriction does not give 
any essential differences w.r.t our main results. 
By the same reason we restrict ID-links to them with literal conclusions. 
Moreover we do not consider Cut-links and Cut-elimination because 
our main results do not concern them. 

\begin{definition}[Literals]
A literal is $p$ or $p^{\bot}$. The positive literal is $p$ and the negative literal is $p^{\bot}$.
\end{definition}

\begin{definition}[MLL Formulas]
MLL formulas (or simply formulas) $F$ is any of the followings:
\begin{itemize}
\item $F$ is a literal; 
\item $F$ is $F_1 \TENS F_2$, where $F_1$ and $F_2$ are MLL formulas. Then $F$ is called $\TENS$-formula. 
\item $F$ is $F_1 \PAR F_2$, where $F_1$ and $F_2$ are MLL formulas. Then $F$ is called $\PAR$-formula. 
\end{itemize}
\end{definition}

We denote the set of all the MLL formulas by $\MLLFml$. 

\begin{definition}[Negations of MLL Formulas]
Let $F$ be an MLL formula. 
The negation $F^{\bot}$ of $F$ is defined as follows according to the form of $F$:
\begin{itemize}
\item if $F$ is $p$, then $F^{\bot} \equiv_{\mathdef} p^{\bot}$;
\item if $F$ is $p^{\bot}$, then $F^{\bot} \equiv_{\mathdef} p$;
\item if $F$ is $F_1 \TENS F_2$, then 
      $F^{\bot} \equiv_{\mathdef} F_1^{\bot} \PAR F_2^{\bot}$;
\item if $F$ is $F_1 \PAR F_2$, then 
      $F^{\bot} \equiv_{\mathdef} F_1^{\bot} \TENS F_2^{\bot}$.
\end{itemize}
\end{definition}
So, $F^{\bot}$ is actually an MLL formula.

\begin{definition}[Indexed MLL Formulas]
An indexed MLL formula is a pair $\langle F, i \rangle$, where
$F$ is an MLL formula and $i$ is a natural number. 
\end{definition}

\begin{figure}[htbp]
\begin{center}
\includegraphics[scale=.5]{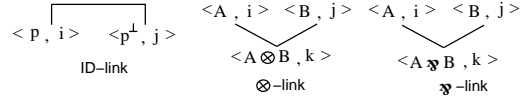}
\caption[MLL links]{MLL links}  
\label{mlllink}
\end{center}
\end{figure}

Figure~\ref{mlllink} shows the links we use in this paper. 
We call each link in Figure~\ref{mlllink} {\it an MLL link} (or simply {\it link}).
In Figure~\ref{mlllink}, 
\begin{enumerate}
\item In ID-link, $\langle A, i \rangle$ and $\langle A^{\bot}, j \rangle $ are called conclusions of the link.
\item In $\TENS$-link (resp. $\PAR$-link)  $\langle A, i \rangle$ is called the left premise, 
 $\langle B, j \rangle$ the right premise 
 and $\langle A \TENS B, k \rangle$ (resp. $\langle A \PAR B, k \rangle$) the conclusion of the link.
\end{enumerate}
Moreover we call links except ID-links {\it multiplicative links}. 

\begin{definition}[MLL Proof Structures]
\label{defProofStructures}
Let ${\mathbb F}$ be a finite set of MLL formula occurrences, i.e., a finite set of indexed MLL formulas  and 
${\mathbb L}$ be a finite set of MLL link occurrences such that 
for each $L \in {\mathbb L}$, the conclusions and the premises of $L$ belong to ${\mathbb F}$. 
The pair $\Theta = \langle {\mathbb F}, {\mathbb L} \rangle$ is 
an MLL proof structure (or simply, a proof structure) if $\Theta$ satisfies the following conditions:
\begin{enumerate}
\item for any $\langle F_0, i \rangle$ and $\langle F'_0, j \rangle$ in ${\mathbb F}$, if $i = j$, then $F_0 = F'_0$ 
(i.e., in ${\mathbb F}$, each element has a different index number).
\item for each formula occurrence $F \in {\mathbb F}$ and for each link occurrence $L \in {\mathbb L}$, 
if $F$ is a premise of $L$ then 
$L$ is unique, i.e., 
$F$ is not a premise of any other link $L' \in {\mathbb L}$.
\item for each formula occurrence $F \in {\mathbb F}$, 
there is a unique link occurrence $L \in {\mathbb L}$ 
such that $F$ is a conclusion of $L$. 
\end{enumerate}
\end{definition}

\begin{figure}[htbp]
\begin{center}
\includegraphics[scale=.5]{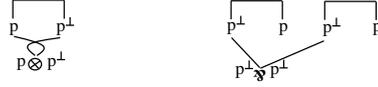}
\caption[Two examples of MLL proof structures]{Two examples of MLL proof structures}  
\label{ps-ex1}
\end{center}
\end{figure}

\begin{remark}
In the following, when we discuss proof structures or proof nets, in many cases, we conveniently forget indices for them, 
because such information is superfluous in many cases. Moreover, when we draw a proof structure or a proof net, 
we also forget such an index, because locative information in such drawings plays an index. 
\end{remark}
We say that in $\Theta = \langle {\mathbb F}, {\mathbb L} \rangle$, 
a formula occurrence $F \in {\mathbb F}$ is a conclusion of $\Theta$ if 
for any $L \in {\mathbb L}$, $F$ is not a premise of $L$. \\
It is well-known that a proof structure does not necessarily correspond to a 
sequent calculus proof. For example, two MLL proof structures 
in Figure~\ref{ps-ex1} do not the corresponding sequent calculus proofs.
The following sequentializability is a judgement 
on the correspondence. 

\begin{definition}[Sequentializability]
A MLL proof structure $\Theta=\langle {\mathbb F}, {\mathbb L} \rangle$ 
is sequentializable if 
any of the following conditions holds:
\begin{enumerate}
\item ${\mathbb L} = \{ L \}$ and $L$ is an ID-link; 
\item There is a $\PAR$-link $L \in {\mathbb L}$ such that
 the conclusion $A \PAR B$ of $L$ is a conclusion of $\Theta$ 
 and $\langle {\mathbb F} - \{ A \PAR B \} , 
	      {\mathbb L} - \{ L \} \rangle$ is sequentializable.
\item There is a $\TENS$-link $L \in {\mathbb L}$ and 
 there are two subsets ${\mathbb F_1}$ and 
 ${\mathbb F_2}$ of ${\mathbb F}$ and two subsets  
 ${\mathbb L_1}$ and ${\mathbb L_2}$ of ${\mathbb L}$ such that
 (a) the conclusion $A \TENS B$ of $L$ is a conclusion of $\Theta$, 
 (b) ${\mathbb F} = {\mathbb F_1} \uplus {\mathbb F_2} 
		     \uplus \{ A \TENS B \}$, 
 (c) ${\mathbb L} = {\mathbb L_1} \uplus {\mathbb L_2} 
		     \uplus \{ L \}$, and 
 (d) $\langle {\mathbb F_1}, {\mathbb L_1} \rangle $ 
 (respectively $\langle {\mathbb F_2}, {\mathbb L_2} \rangle$) is 
 an MLL proof structure and sequentializable, 
 where $\uplus$ denotes the disjoint union operator. 

\end{enumerate}
\end{definition}

\begin{figure}[htbp]
\begin{center}
\includegraphics[scale=.5]{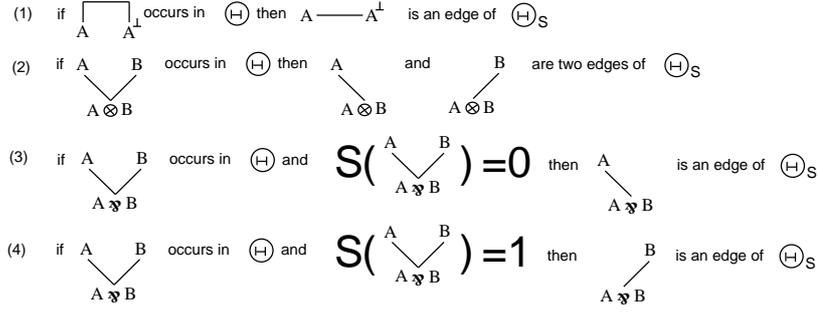}
\caption[The rules for the generation of the edges of a Danos-Regnier graph $\Theta_S$]{The rules for the generation of the edges of a Danos-Regnier graph $\Theta_S$}  
\label{dr-graph-edges}
\end{center}
\end{figure}

\begin{definition}[MLL Proof Nets]
An MLL proof structure $\Theta$ is an MLL proof net if 
$\Theta$ is sequentializable.  
\end{definition}
Next we give a graph-theoretic characterization of MLL proof nets, 
following \cite{Gir96}.
The characterization was firstly proved in \cite{Gir87} and 
then an improvement was given in \cite{DR89}. 
In order to characterize MLL proof nets among MLL proof structures, 
we introduce {\it Danos-Regnier graphs (for short, DR-graphs)}. 
Let $\Theta$ be an MLL proof structure. 
We assume that we are given a function $S$ from the set of the occurrences of
$\PAR$-links in $\Theta$ to $\{ 0, 1 \}$. 
Such a function is called a {\it DR-switching} for $\Theta$.
Then the Danos-Regnier graph $\Theta_S$ for $\Theta$ and $S$ is a 
undirected graph such that
\begin{enumerate}
\item the nodes are all the formula occurrences in $\Theta$, and 
\item the edges are generated by the rules of Figure~\ref{dr-graph-edges}.
\end{enumerate}
In the following we also use the alternative notation $S(\Theta)$ for the Danos-Regnier graph $\Theta_S$. \\
The following theorem by Girard, Danos, and Regnier \cite{Gir87,DR89}, 
which is called {\it sequentialization theorem},  is the most important theorem in the theory of MLL proof nets. 
\begin{theorem}
\label{seqthm}
An MLL proof structure $\Theta$ is an MLL proof net 
iff 
for each switching function $S$ for $\Theta$, 
the Danos-Regnier graph $\Theta_S$ is acyclic and connected.
\end{theorem}

\subsection{Empires}
In this subsection, we introduce {\it empires} following \cite{Gir06}.
The notion is needed to establish our main results.
First we fix a proof structure $\Theta = \langle {\mathbb F}_{\Theta}, {\mathbb L}_{\Theta} \rangle$. 
Moreover we introduce the notations $\fml(\Theta) \equiv_{\mathdef} {\mathbb F}_{\Theta}$ and 
$\lnk(\Theta) \equiv_{\mathdef} {\mathbb L}_{\Theta}$. 

\begin{definition}[Empires]
The empire of a formula $A$ in 
a proof net $\Theta = \langle {\mathbb F}, {\mathbb L} \rangle$ (denoted by $e_\Theta(A)$) is defined in 
the following manner: let $S$ be a DR-switching for $\Theta$. 
Then an undirected maximal connected graph ${(\Theta_S)}^{A}$ (or simply ${\Theta_S}^A$) is defined as follows:
\begin{enumerate}
\item If there is a link $L \in E$ such that $A$ is a premise of $L$
      and there is the edge $e$ from $A$ to the conclusion of $L$ in $\Theta_S$, then 
      ${(\Theta_S)}^{A}$ is the maximal connected graph including $A$ obtained from  $\Theta_S$ by deleting $e$;
\item otherwise, ${(\Theta_S)}^{A} = \Theta_S$.
\end{enumerate}
Then the empire $A$ in $\Theta$ (denoted by $e_\Theta(A)$) is defined as follows:
\[ e_\Theta(A) \equiv_{\mathdef} 
\bigcap_{S \, \, \mbox{\scriptsize is a DR-switching for} \, \, \Theta} \fml({\Theta_S}^{A}) \]
\end{definition}

From the definition it is obvious that $A \in e_\Theta(A)$. 
Although the empire $e_\Theta(A)$ is defined as a set of formula occurrences, 
by considering the set ${\mathbb L}_{e_\Theta(A)}$of links whose conclusions and premises are all included 
in $e_\Theta(A)$, 
the empire $e_\Theta(A)$ can be considered as the pair 
$\langle e_\Theta(A), {\mathbb L}_{e_\Theta(A)} \rangle$. \\
Appendix~\ref{secEmpBasicProp} gives basic properties on empires. Many of them are used in Section~\ref{secPSFamilies}.

\section{Families of Proof Structures} \label{secPSFamilies}
\subsection{Our Framework}
\label{secOurFramework}
Firstly we define families of proof-structures. 
Informally two proof structures $\Theta_1$ and $\Theta_2$ that belong to the same family means that 
$\Theta_2$ is obtained from $\Theta_1$ by replacing several $\TENS$-links (resp. $\PAR$-links) 
by $\PAR$-links (resp. $\TENS$-links). 
We define such families using graph isomorphisms on directed graphs in a mathematically rigorous way.
The reader might feel that the following definitions in this subsection are too cumbersome. 
But there is a subtle point of the definitions. 
That is the reason why we insist on a rigorous style. 
We will discuss this matter at the end of the subsection. 

\begin{definition}[Strip Function]
\label{defStrip}
A function $\stripPT: \MLLFml \to \{ p, p^{\bot}, \TENS, \PAR \}$ 
is defined
as follows:
\begin{enumerate}
\item $\stripPT(p) = p$ and $\stripPT(p^{\bot}) = p^{\bot}$;
\item $\stripPT(A \TENS B) = A \TENS B$ and $\stripPT(A \PAR B) = \PAR$. 
\end{enumerate} 
\end{definition}

\begin{definition}[Labelled Directed Graphs]
Let ${\mathbb A}$ and ${\mathbb B}$ be sets. 
A labelled directed graph with labels ${\mathbb B}$ (resp. ${\mathbb A}$ and ${\mathbb B}$) 
is a tuple $\langle V, E, \ell_E: E \to {\mathbb B} \rangle$ 
(resp. $\langle V, E, \ell_V: V \to {\mathbb A}, \ell_E: E \to {\mathbb B} \rangle$)
 satisfying the following conditions:
\begin{enumerate} 
\item [1.] $V$ is a set;
\item [2.] $E$ is a set with two functions 
    $\src : E \to V$ and $\tgt : E \to V$.
\end{enumerate}
\end{definition}
In the following, we suppose ${\mathbb A} = \{ p, p^{\bot}, \TENS, \PAR \}$
and 
${\mathbb B} = \{ {\bf L}, {\bf R}, {\bf ID} \}$. 

Next we define a translation from proof structures to labelled directed graphs 
and that with a function $f : \MLLFml \to {\mathbb A}$ as a parameter. 

 \begin{definition}[Labelled Directed Graphs Induced by Proof Structures]
\label{deflabelledgraphG}
 Let $\Theta = \langle {\mathbb F}, {\mathbb L} \rangle$ be a proof structure
 and $f : \MLLFml \to {\mathbb A}$. 
 A labelled directed graph 
 $G(\Theta) = \langle V, E, \ell_E : E \to \{ {\bf L}, {\bf R}, {\bf ID} \} \rangle$ 
 and
 $G^f(\Theta) = \langle V, E, \ell_V^{f} : V \to {\mathbb A}, \ell_E : E \to \{ {\bf L}, {\bf R}, {\bf ID} \} \rangle$ 
 is defined from $\Theta$ in the following way:
 \begin{enumerate}
 \item $V = \{ i \, | \, \langle A, i \rangle \in {\mathbb F} \}$ and 
   $\ell_V^{f} = \{ \langle i, f(A) \rangle \, | \, \langle A, i \rangle \in {\mathbb F} \}$;\\
   Since in $\Theta$, each formula occurrence has a unique index, we can easily see that 
   $V$ is bijective to ${\mathbb F}$.
 \item $E$ and $\ell_E$ is the least set satisfying the following conditions: 
   \begin{itemize}
     \item If $L \in {\mathbb L}$ is an ID-link occurrence with conclusions $\langle p, i \rangle$ and $\langle p^{\bot}, j \rangle$, then 
	there is an edge $e \in E$ such that $\src(e)=i$ and $\tgt(e)=j$	
	and $\langle e, {\bf ID} \rangle \in \ell_E$;
     \item If $L \in {\mathbb L}$ is a $\TENS$-link occurrence with the form
       $\frac{\langle A, i \rangle \quad \langle B, j \rangle}{\langle A \TENS B, k \rangle}$ , then 
	 there are two edges $e_1 \in E$ and  $e_2 \in E$ such that
	 $\src(e_1)=i$, $\tgt(e_1)=k$, $\src(e_2)=j$, $\tgt(e_2)=k$, 
       $\langle e_1, {\bf L} \rangle \in \ell_E$, and  $\langle e_2, {\bf R} \rangle \in \ell_E$; 
     \item If $L \in {\mathbb L}$ is a $\PAR$-link occurrence with the form
       $\frac{\langle A, i \rangle \quad \langle B, j \rangle}{\langle A \PAR B, k \rangle}$ , then
       there are two edges $e_1 \in E$ and $e_2 \in E$ such that
       $\src(e_1)=i$, $\tgt(e_1)=k$, $\src(e_2)=j$, $\tgt(e_2)=k$, 
       $\langle e_1, {\bf L} \rangle \in \ell_E$,  and  $\langle e_2, {\bf R} \rangle \in \ell_E$.
   \end{itemize}
 \end{enumerate}
 \end{definition}

The next definition is a slight extension of the standard definition of graph isomorphisms. 

\begin{definition}[Graph Isomorphisms on Labelled Directed Graphs]
\label{defgiso}
Let \\
$G_1 = \langle V_1, E_1, \ell_{E_1} \rangle$ 
(resp. $G_1 = \langle V_1, E_1, \ell_{V_1}, \ell_{E_1} \rangle$) and 
$G_2 = \langle V_2, E_2, \ell_{V_2}, \ell_{E_2} \rangle$
(resp. $G_2 = \langle V_2, E_2, \ell_{E_2} \rangle$) be labelled directed graphs. 
Then a graph homomorphism from $G_1$ to $G_2$ is a pair $\langle h_V: V_1 \to V_2, h_E : E_1 \to E_2 \rangle$ 
satisfying the following conditions:
\begin{enumerate}
\item [1.] for any $e \in E_1$, $h_V(\src(e)) = \src(h_E(e))$ and $h_V(\tgt(e)) = \tgt(h_E(e))$;
\item [2.] (only the case where $\ell_{V_1}$ and $\ell_{V_2}$ are specified) for any $v \in V_1$, $\ell_{V_1}(v) = \ell_{V_2}(h_V(v))$;
\item [3.] for any $e \in E_1$, $\ell_{E_1}(e) = \ell_{E_2}(h_E(e))$.
\end{enumerate}
The graph homomorphism $\langle h_V, h_E \rangle$ is a graph isomorphism if 
$h_V : V_1 \to V_2$ and $h_E : E_1 \to E_2$ are both bijections 
(then, we write $\langle h_V, h_E \rangle : G_1 \simeq G_2$). 

\end{definition}
\begin{definition}[PS-families]
Let $\Theta_1$ and $\Theta_2$ be proof structures. 
Then $\Theta_1 \sim \Theta_2$ if 
there is a graph isomorphism $\langle h_V : V_1 \to V_2, h_E : E_1 \to E_2 \rangle$ from 
$G(\Theta_1) = \langle V_1, E_1, \ell_{E_1} \rangle$ to 
$G(\Theta_2) = \langle V_2, E_2, \ell_{E_2} \rangle$.
It is obvious that $\sim$ is an equivalence relation. 
Therefore for a given proof structure $\Theta$, we can define the equivalence class $[\Theta]$ such that 
$\Theta' \in [\Theta]$ iff $\Theta \sim \Theta'$. 
Then we say $[\Theta]$ is a {\bf PS-family} of $\Theta$. We also say $\Theta$ belongs to the PS-family $[\Theta]$. 
\end{definition}

\begin{remark}
\label{remPSFamilies}
We define a PS-family as an equivalence class generated by the relation $\sim$.
Of course, we can define a PS-family as an MLL proof structure 
in which all the occurrences of multiplicative links are of $\frac{A \quad B}{A @ B}$ 
instead of $\TENS$- and $\PAR$-links, where $@$ is a new symbol. The reader might prefer to this form. 
But it seems a matter of taste. 
\end{remark}
We denote a PS-family by ${\mathcal F}$. \\
Next, given a PS-family ${\mathcal F}$, we introduce a metric $d_{\mathcal F}$ on ${\mathcal F}$. 
\begin{definition}
\label{defMetricSpaceonPSFamilies}
Let ${\mathcal F}$ be a PS-family.
We assume that two MLL proof structures $\Theta_1$ and $\Theta_2$ belong to ${\mathcal F}$. 
So, by definition we have at least one graph isomorphism $\langle h_V, h_E \rangle$ 
from $G(\Theta_1)$ to $G(\Theta_2)$. 
Moreover let 
$G^{\stripPT}(\Theta_1) = \langle V_1, E_1, \ell_{V_1}^{\stripPT}, \ell_{E_1} \rangle$ and 
$G^{\stripPT}(\Theta_2) = \langle V_2, E_2, \ell_{V_2}^{\stripPT}, \ell_{E_2} \rangle$. 
Then $d_{\mathcal F}(\Theta_1, \Theta_2) \in \mathbb{N}$ is defined as follows:
\[ d_{\mathcal F}(\Theta_1, \Theta_2) = 
\min \{ | \{ v_1 \in V_1 \, | \,  \ell_{V_2}^{\stripPT}(h_V(v_1)) \neq \ell_{V_1}^{\stripPT}(v_1) \} | \, \, \, | \, \langle h_V, h_E \rangle : G(\Theta_1) \simeq G(\Theta_2) \} \]
\end{definition}
Before proving that $\langle {\mathcal F}, d_{\mathcal F} \rangle$ is a metric space, 
we must define an equality between two MLL proof structures,
because the statement concerns the equality on ${\mathcal F}$. 
In order to define the equality, we use Definition~\ref{deflabelledgraphG} with the parameter $\stripPT$. 

\begin{definition}[Equality on MLL Proof Structures]
\label{defEqMLLPS}
Let $\Theta_1$ and $\Theta_2$ be proof structures. 
Then $\Theta_1 = \Theta_2$ if 
there is a graph isomorphism $\langle h_V : V_1 \to V_2, h_E : E_1 \to E_2 \rangle$ from 
$G^{\stripPT}(\Theta_1) = \langle V_1, E_1, \ell_{V_1}^{\stripPT}, \ell_{E_1} \rangle$ to 
$G^{\stripPT}(\Theta_2) = \langle V_2, E_2, \ell_{V_2}^{\stripPT}, \ell_{E_2} \rangle$.
\end{definition}
It is obvious that $=$ is an equivalence relation. \\

\begin{proposition}
\label{propDMetric}
The pair $\langle {\mathcal F}, d_{\mathcal F} : {\mathcal F} \to \mathbb{N} \rangle$ is a metric space.
\end{proposition}

\begin{figure}[htbp]
\begin{center}
\includegraphics[scale=.5]{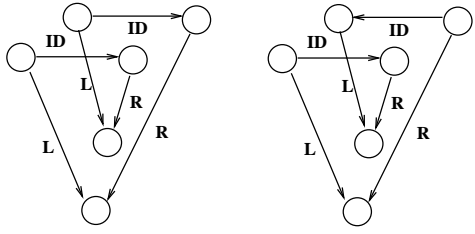}
\caption[The induced directed graphs from $\Theta_1$ and $\Theta_2$, and that of $\Theta_3$]
{The induced directed graphs from $\Theta_1$ and $\Theta_2$, and that of $\Theta_3$}  
\label{inducedDirectedGraph}
\end{center}
\end{figure}

\begin{figure}[htbp]
\begin{center}
\includegraphics[scale=.5]{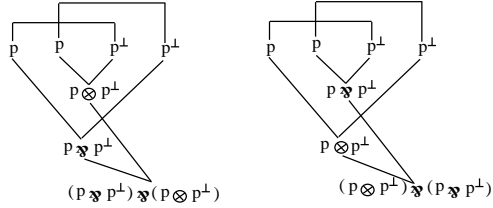}
\caption[Two elements of a closed PS-family]
{Two elements of a closed PS-family}  
\label{explanatoryExampleClosed}
\end{center}
\end{figure}

\begin{proof}
The non-negativity of $d_{\mathcal F}$ is also obvious. 
It is obvious that $d_{\mathcal F}$ is symmetry. \\
The formula $\Theta_1 = \Theta_2 \Rightarrow d_{\mathcal F}(\Theta_1, \Theta_2)=0$ is obvious. 
Next we prove that $d_{\mathcal F}(\Theta_1, \Theta_2) = 0 \Rightarrow \Theta_1 = \Theta_2$. 
Let 
$G(\Theta_1) = \langle V_1, E_1, \ell_{E_1} \rangle$ and
$G(\Theta_2) = \langle V_2, E_2, \ell_{E_2} \rangle$. 
Since $\Theta_1$ and $\Theta_2$ belong to the same PS-family ${\mathcal F}$, 
there is a graph isomorphism $\langle h_V: V_1 \to V_2, h_E: E_1 \to E_2 \rangle$
from $G(\Theta_1)$ to $G(\Theta_2)$. 
By Definition~\ref{defgiso}, this means that
both $h_V: V_1 \to V_2$ and $h_E : E_1 \to E_2$ are bijections and 
\begin{enumerate}
\item [1.] for any $e \in E_1$, $h_V(\src(e)) = \src(h_E(e))$ and $h_V(\tgt(e)) = \tgt(h_E(e))$;
\item [2.] for any $e \in E_1$, $\ell_{E_1}(e) = \ell_{E_2}(h_E(e))$.
\end{enumerate}
On the other hand, since $d_{\mathcal F}(\Theta_1, \Theta_2) = 0$, 
we find a graph isomorphism $\langle h_V: V_1 \to V_2, h_E: E_1 \to E_2 \rangle : G(\Theta_1) \to G(\Theta_2)$
with the following additional property: 
for any $v \in V_1$, $\ell_{V_1}^{\stripPT}(v) = \ell_{V_2}^{\stripPT}(h_V(v))$. 
So, we have a graph isomorphism from $G^{\stripPT}(\Theta_1)$ to $G^{\stripPT}(\Theta_2)$. 
By Definition~\ref{defEqMLLPS}, we obtain $\Theta_1 = \Theta_2$. \\
In order to prove the triangle equality on $d_{\mathcal F}$, we need the following claim. 
\begin{claim}
\label{claimPSdistanceSubset}
Let $\Theta_1$ and $\Theta_2$ be two proof structures belonging to the same PS-family $\mathbb{F}$. 
Moreover let $\langle h_V, h_E \rangle : G(\Theta_1) \simeq G(\Theta_2)$,  
$V_h = \{ v_1 \in V_1 \, | \,  \ell_{V_2}^{\stripPT}(h_V(v_1)) \neq \ell_{V_1}^{\stripPT}(v_1) \}$, 
and 
$d_{\mathcal F}(\Theta_1, \Theta_2) = | V_h |$. 
In addition let $V' \subseteq V_h$ and $\Theta_0$ be the proof structure obtained from 
$\Theta_1$ by replacing the $\TENS$-link (resp. the $\PAR$-link) corresponding to $v$ by 
the $\PAR$-link (resp. the $\TENS$-link) for each $v \in V'$. 
Then $d_{\mathcal F}(\Theta_1, \Theta_0) = |V'|$. 
\end{claim}
{\it proof of Claim~\ref{claimPSdistanceSubset}: \ }
We assume that $d_{\mathcal F}(\Theta_1, \Theta_0) < |V'|$. 
Then we have $\langle h_{V}^{0}, h_{E}^{0} \rangle : G(\Theta_1) \simeq G(\Theta_0)$ such that 
$d_{\mathcal F}(\Theta_1, \Theta_0) = 
| \{ v_1 \in V_1 \, | \,  \ell_{V_0}^{\stripPT}(h_{V}^{0}(v_1)) \neq \ell_{V_1}^{\stripPT}(v_1) \} |$. \\
On the other hand, 
$\langle h_V, h_E \rangle : G(\Theta_1) \simeq G(\Theta_2)$ can be decomposed into 
$\langle h_{V}^{10}, h_{E}^{10} \rangle : G(\Theta_1) \simeq G(\Theta_0)$ and 
$\langle h_{V}^{02}, h_{E}^{02} \rangle : G(\Theta_0) \simeq G(\Theta_2)$ 
(i.e., $\langle h_V, h_E \rangle = \langle h_{V}^{02}, h_{E}^{02} \rangle \circ \langle h_{V}^{10}, h_{E}^{10} \rangle$) 
such that 
$| V_{10} | + | V_{02} | = |V_h|$, 
where
$V_{10} = \{ v_1 \in V_1 \, | \,  \ell_{V_0}^{\stripPT}(h_{V}^{10}(v_1)) \neq \ell_{V_1}^{\stripPT}(v_1) \}$
and $V_{02} = \{ v_0 \in V_0 \, | \,  \ell_{V_2}^{\stripPT}(h_{V}^{02}(v_0)) \neq \ell_{V_0}^{\stripPT}(v_0) \}$. 
We note $|V'| = |V_{10}|$. 
Then $\langle h_{V}^{02} \circ h_{V}^{0}, h_{E}^{02} \circ h_{E}^{0} \rangle : G(\Theta_0) \simeq G(\Theta_2)$ and
\begin{eqnarray*} 
& & | \{ v_1 \in V_1 \, | \,  \ell_{V_2}^{\stripPT}((h_{V}^{02} \circ h_{V}^{0})(v_1)) \neq \ell_{V_1}^{\stripPT}(v_1) \} | \\
& = & d_{\mathcal F}(\Theta_1, \Theta_0) + |V_{20}| < |V'| + |V_{02}| = |V_{10}| + |V_{02}| = 
|V_h| = d_{\mathcal F}(\Theta_1, \Theta_2) 
\, \,. 
\end{eqnarray*}
This is a contradiction. 
\ {\it the end of the proof of Claim~\ref{claimPSdistanceSubset} \ } \\
Using the claim, we can prove the triangle equality on $d_{\mathcal F}$ similar to 
that of the set of all the binary words with a fixed length. 
$\Box$
\end{proof}
We give a justification of the definitions above using Figure~\ref{explanatoryEx}. 
Let $\Theta_1$, $\Theta_2$, and $\Theta_3$ be the left proof net, the middle proof net, and the right proof net 
of Figure~\ref{explanatoryEx} respectively. 
Then  $G(\Theta_1) \sim G(\Theta_2)$, 
since $G(\Theta_1)$ and $G(\Theta_2)$ are graph-isomorphic to the left directed graph of 
Figure~\ref{inducedDirectedGraph}. 
But note that there are two graph isomorphisms $\{ \TENS \mapsto \TENS, \PAR \mapsto \PAR \}$ 
and $\{ \TENS \mapsto \PAR, \PAR \mapsto \TENS \}$ between $G(\Theta_1)$ and $G(\Theta_2)$. 
By the former one, we can identify $\Theta_1$ with $\Theta_2$, 
while in the latter one, there are two differences w.r.t multiplicative nodes. 
Therefore $d_{\mathbb{F}}(\Theta_1, \Theta_2) = 0$. 
That's why we need the $\min$ operator for the definition of $d_{\mathbb{F}}(\Theta_1, \Theta_2)$. 
So, $\Theta_1$ and $\Theta_2$ belong to the same PS-family. 
But $\neg (G(\Theta_1) \sim G(\Theta_3))$ 
(and also $\neg (G(\Theta_2) \sim G(\Theta_3))$), 
since $G(\Theta_3)$ is graph-isomorphic to the right directed graph of Figure~\ref{inducedDirectedGraph}
and the left one of Figure~\ref{inducedDirectedGraph} are not graph-isomorphic to the right one. 
So, $\Theta_3$ does not belong to the same PS-family as $\Theta_1$ and $\Theta_2$. \\
Note that direction of edges labelled with ${\bf ID}$ are indispensable, because 
if we eliminated the information, then the two graphs of Figure~\ref{inducedDirectedGraph} would be isomorphic. 
However, direction of edges labelled with ${\bf L}$ or ${\bf R}$ is redundant, 
because we can always identify the conclusions of the graph without the information 
by looking for the nodes without an outgoing edge. 
But we prefer to the conventional definition of directed graphs.  \\
In order to avoid the $\min$ operator for the definition of $d_{\mathbb{F}}(\Theta_1, \Theta_2)$, 
we need to consider only PS-families in which 
there is the unique graph isomorphism between $G(\Theta_1)$ and $G(\Theta_2)$
for each two members $\Theta_1$ and $\Theta_2$. 
In order to do that, we restrict PS-families to them with exactly one conclusion, 
because each multiplicative link in an element in such a PS-family is given an absolute position 
from the root of the proof structure. 
We call such a PS-family {\it closed PS-family}. 
A closed PS-family is PS-connected in the sense of Definition~\ref{defPSConnected} (Subsection~\ref{subsecOtherTopics}). 
For example, two proof structures in Figure~\ref{explanatoryExampleClosed} belonging to the same closed PS-family
has the unique graph isomorphism between them. 
The restriction is similar to that of closed loops in knot theory (see \cite{Ada94}). \\
On the other hand, for any MLL proof net without closedness condition, the following proposition holds.
\begin{proposition}
\label{propPNGraphAutoUnique}
Let $\Theta$ be an MLL proof net.
Then the identity map $\langle \id_V, \id_E \rangle$ is 
the only one graph automorphism on 
$G^{\stripPT}(\Theta) = \langle V, E, \ell_{V}^{\stripPT}, \ell_{E} \rangle$.
\end{proposition}
Our proof of Proposition~\ref{propPNGraphAutoUnique} is given in Appendix~\ref{secPNGraphAutoUnique}.

\subsection{Basic Results}
Our proposal in this paper starts from the following trivial proposition.
We note that this proposition is stated in Subsection 11.3.3 of \cite{Gir06}.
\begin{proposition}
\label{propMainKeyObserve}
Let $\Theta$ be an MLL proof net. 
\begin{enumerate}
\item [1.] Let $L_{\TENS} : \frac{A \quad B}{A \TENS B}$ 
be a $\TENS$-link in $\Theta$. 
Let $\Theta'$ be the proof structure $\Theta$ except that 
$L_{\TENS}$ is replaced by 
$L'_{\PAR} : \frac{A \quad B}{A \PAR B}$. 
Then $\Theta'$ is not an MLL proof net. 
\item [2.] Let $L_{\PAR} : \frac{C \quad D}{C \PAR D}$ 
be a $\PAR$-link in $\Theta$. 
Let $\Theta''$ be the proof structure $\Theta$ except that 
$L_{\PAR}$ is replaced by 
$L'_{\TENS} : \frac{C \quad D}{C \TENS D}$. 
Then $\Theta'$ is not an MLL proof net. 
\end{enumerate}
\end{proposition}

\begin{proof}
\begin{enumerate}
\item [1.] It is obvious that 
    there is a formula $X$ (resp. $Y$) in $\fml(\Theta)$ such that 
    $X \neq A$ (resp. $Y \neq B$) and 
    $X \in e_{\Theta}(A)$ 
    (resp. $Y \in e_{\Theta}(B)$) since 
    if $A$ (resp. $B$) 
    is a literal, then we just take $X$ (resp. $Y$) as the other conclusion of the ID-link 
    whose conclusion is $A$ (resp. $B$), 
    and otherwise, we just take $X$ (resp. $Y$) as the formula immediately above $A$ (resp. $B$). 
    On the other hand since 
    $e_\Theta(A) \cap e_\Theta(B) = \emptyset$ 
    by Proposition~\ref{propEmpireTensPremise}, when we pick up a DR-switching $S$ for $\Theta$ arbitrarily, 
    the unique path $X$ from $Y$ in $S(\Theta)$ always passes 
    $A, A \TENS B, B$. 
    Then let $S'$ be a DR-switching for $\Theta'$ obtained from $S$ by adding a selection for 
    $L'_{\PAR}$. Then it is obvious that $X$ and $Y$ is disconnected in $S'(\Theta')$. 
\item [2.] Let $S$ be a DR-switching for $\Theta$. Then by Proposition~\ref{propEmpireParPremise} 
      there is the unique path $\theta$ from $C$ to $D$ in $S(\Theta)$
      such that $\theta$ does not include $C \PAR D$. 
      Then Let $S''$ be the DR-switching for $\Theta''$  obtained form $S$ by deleting 
      the $\PAR$-switch for $L_{\PAR}$. It is obvious that $S''(\Theta'')$ has a cycle including $\theta$ and 
      $C \TENS D$. 
      $\Box$
\end{enumerate}
\end{proof}

\begin{remark}
\label{remTensParExchange}
Proposition~\ref{propMainKeyObserve} does not hold in neither MLL+MIX \cite{Gir87} nor Affine Logic \cite{Bla92}.
For example $(p \PAR p^{\bot}) \TENS (p \PAR p^{\bot})$ is provable in MLL, MLL+MIX, and Affine Logic.
The formula $(p \PAR p^{\bot}) \PAR (p \PAR p^{\bot})$ is not provable in MLL, 
but provable in both MLL+MIX and Affine Logic, 
\end{remark}

The following corollary is obvious.
\begin{corollary}
\label{corOneErrorDetect}
Let $\Theta_1$ and $\Theta_2$ be MLL proof nets belonging to the same PS-family ${\mathcal F}$. 
Then $d_{\mathcal F}(\Theta_1, \Theta_2) \ge 2$. 
\end{corollary}

This corollary says that if a PS-family ${\mathcal F}$ has $n$ MLL proof nets, then 
${\mathcal F}$ can be used as a {\bf one error-detecting code system} with 
$n$ different code elements(see Appendix~\ref{secBasicCodeTheory}). 
But since neither MLL+MIX nor Affine Logic has the property, 
these can not be used as such a system. \\
The following proposition is basically a slight extension of Corollary 17.1 of Subsection 11.A.2 of \cite{Gir06}.
The extension is by a suggestion of an anonymous referee of the previous version of this paper. 
\begin{proposition}
\label{propNumAxAndTensCons}
Let $\Theta = \langle {\mathbb F}_\Theta, {\mathbb L}_\Theta \rangle$ be an MLL proof net. 
Let ${\mathbb L}_{\Theta}^{\ID}$, ${\mathbb L}_{\Theta}^{\TENS}$, and ${\mathbb L}_{\Theta}^{\PAR}$
be 
the set of the ID-links, the $\TENS$-links, and the $\PAR$-links in ${\mathbb L}$ respectively and 
${\con}_{\Theta}$ be the set of the conclusions in ${\mathbb F}_\Theta$. 
Then $|{\con}_{\Theta}| + |{\mathbb L}_{\Theta}^{\PAR}| = |{\mathbb L}_{\Theta}^{\ID}| + 1$
and 
$|{\mathbb L}_{\Theta}^{\ID}| - |{\mathbb L}_{\Theta}^{\TENS}| = 1$. 
\end{proposition}

\begin{proof}
We prove this by induction on $|{\mathbb L}_\Theta|$.
\begin{enumerate}
\item The case where $|{\mathbb L}_\Theta|=1$:\\
   Then $|{\mathbb L}_{\Theta}^{\ID}| = |{\mathbb L}|=1$, $|{\con_{\Theta}}|=2$, and  
   $|{\mathbb L}_{\Theta}^{\PAR}| = |{\mathbb L}_{\Theta}^{\TENS}|=0$. 
   The statements holds obviously. 
\item The case where $|{\mathbb L}_\Theta|>1$:
\begin{enumerate} 
\item The case where $\Theta$ includes a $\PAR$-formula as a conclusion:\\
  We choose one $\PAR$-link $L_{\PAR}$ among such $\PAR$-links. 
  Let $\Theta_0= \langle {\mathbb F}_{\Theta_0}, {\mathbb L}_{\Theta_0} \rangle$ be 
  $\Theta$ except that $L_{\PAR}$ is removed. 
  Since $\Theta_0$ is also an MLL proof net (otherwise, $\Theta$ is not an MLL proof net), 
  by inductive hypothesis 
  $|{\con}_{\Theta_0}| + |{\mathbb L}_{\Theta_0}^{\PAR}| = |{\mathbb L}_{\Theta_0}^{\ID}| + 1$
  and 
  $|{\mathbb L}_{\Theta_0}^{\ID}| - |{\mathbb L}_{\Theta_0}^{\TENS}| = 1$. 
  But since $|{\mathbb L}_{\Theta}^{\ID}|=|{\mathbb L}_{\Theta_0}^{\ID}|$, ${\con}_{\Theta} = {\con}_{\Theta_0}-1$, 
  and $|{\mathbb L}_{\Theta}^{\PAR}| = |{\mathbb L}_{\Theta_0}^{\PAR}|+1$, 
  $|{\mathbb L}_{\Theta}^{\TENS}| = |{\mathbb L}_{\Theta_0}^{\TENS}|$, 
  the statements hold. 
\item The case where the conclusions of $\Theta$ do not have any $\PAR$-formula:\\
  In this case, $|{\mathbb L}_{\Theta}^{\TENS}|$ must be greater than 0. 
  Then by Splitting lemma (Lemma~\ref{lemmaSplitting}), we have a $\TENS$-conclusion $A \TENS B$ and its 
  $\TENS$-link $L_{A \TENS B}$ in $\Theta$ such that 
  $\Theta$ is decomposed into $\Theta_1 = e_{\Theta}(A)$, 
  $\Theta_2 = e_{\Theta}(B)$, and 
  $\TENS$-link $L_{A \TENS B}$ 
  By inductive hypothesis 
  $|{\con}_{\Theta_1}| + |{\mathbb L}_{\Theta_1}^{\PAR}| = |{\mathbb L}_{\Theta_1}^{\ID}| + 1$, 
  $|{\mathbb L}_{\Theta_1}^{\ID}| - |{\mathbb L}_{\Theta_1}^{\TENS}| = 1$, 
  $|{\con}_{\Theta_2}| + |{\mathbb L}_{\Theta_2}^{\PAR}| = |{\mathbb L}_{\Theta_2}^{\ID}| + 1$, 
  and 
  $|{\mathbb L}_{\Theta_2}^{\ID}| - |{\mathbb L}_{\Theta_2}^{\TENS}| = 1$ hold. 
  Moreover since $|{\mathbb L}_{\Theta}^{\ID}| = |{\mathbb L}_{\Theta_1}^{\ID}| + |{\mathbb L}_{\Theta_2}^{\ID}|$, 
  $|{\con}_{\Theta}| = |{\con}_{\Theta_1}| + |{\con}_{\Theta_2}| -1$, 
  $|{\mathbb L}_{\Theta}^{\PAR}| = |{\mathbb L}_{\Theta_1}^{\PAR}| + |{\mathbb L}_{\Theta_2}^{\PAR}|$, 
  and
  $|{\mathbb L}_{\Theta}^{\TENS}| = |{\mathbb L}_{\Theta_1}^{\TENS}| + |{\mathbb L}_{\Theta_2}^{\TENS}| + 1$, 
  the statements holds. 
  $\Box$
\end{enumerate}
\end{enumerate}
\end{proof}

\begin{remark}
\label{remNumAxAndTensCons}
Proposition~\ref{propNumAxAndTensCons} does not hold in MLL+MIX. 
A counterexample in MLL+MIX is again $(p \PAR p^{\bot}) \PAR (p \PAR p^{\bot})$. 
\end{remark}

\begin{corollary}
\label{corPS-familyPNsInvariant}
Let ${\mathcal F}$ be a PS-family.
Let $\Theta_1$ and $\Theta_2$ be MLL proof nets belonging to ${\mathcal F}$. 
Then the number of $\TENS$-links (resp. $\PAR$-links) occurring in $\Theta_1$ 
is the same as that of $\Theta_2$.
\end{corollary}

\begin{proof}
Since $\Theta_1$ and $\Theta_2$ are members of ${\mathcal F}$, 
$|{\con}_{\Theta_1}| = |{\con}_{\Theta_2}|$ and 
$|{\mathbb L}_{\Theta_1}^{\ID}| = |{\mathbb L}_{\Theta_2}^{\ID}|$. 
Therefore by Proposition~\ref{propNumAxAndTensCons}, 
$|{\mathbb L}_{\Theta_1}^{\TENS}| = |{\mathbb L}_{\Theta_2}^{\TENS}|$ and
$|{\mathbb L}_{\Theta_1}^{\PAR}| = |{\mathbb L}_{\Theta_2}^{\PAR}|$.
$\Box$
\end{proof}

\begin{figure}[htbp]
\begin{center}
\includegraphics[scale=.5]{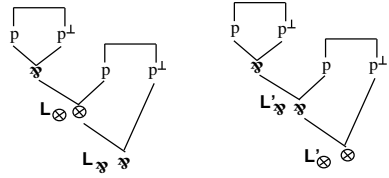}
\caption[A counterexample]{A counterexample}
\label{figTPExchangeCounterExample}
\end{center}
\end{figure}

Next, we define an important notion in the next subsection. 
\begin{definition}[$\TENS$-$\PAR$-exchange]
\label{defTPExchange}
Let $\Theta$ be a proof structure.
Moreover let $L_{\TENS}: \frac{A \quad B}{A \TENS B}$ and 
$L_{\PAR}: \frac{C \quad D}{C \PAR D}$ be 
a $\TENS$-link and a $\PAR$-link in $\Theta$ respectively.
Then $\tpex(\Theta, L_{\TENS}, L_{\PAR})$ be a proof structure obtained from $\Theta$ replacing 
$L_{\TENS}$ by 
$L'_{\PAR}: \frac{A \quad B}{A \PAR B}$ and
$L_{\PAR}$ by
$L'_{\TENS}: \frac{C \quad D}{C \TENS D}$ 
simultaneously. 
Then $\tpex(\Theta, L_{\TENS}, L_{\PAR})$ is called a $\TENS$-$\PAR$-exchange of $\Theta$ by 
$L_{\TENS}$ and $L_{\PAR}$. \\
More generally, when $\langle L_{\TENS_1}, \ldots, L_{\TENS_{\ell_1}} \rangle$ is a list of $\TENS$-links 
and $\langle L_{\PAR_1}, \ldots, L_{\PAR_{\ell_2}} \rangle$ a list of $\PAR$-links, then \\
$\tpex(\Theta, \langle L_{\TENS_1}, \ldots, L_{\TENS_{\ell_1}} \rangle, \langle L_{\PAR_1}, \ldots, L_{\PAR_{\ell_2}} \rangle)$ is defined to be a proof structure obtained from $\Theta$ by 
replacing $L_{\TENS_1}, \ldots, L_{\TENS_{\ell_1}}$ by the list of $\PAR$-links 
$L'_{\PAR_1}, \ldots, L'_{\PAR_{\ell_1}}$ and 
$L_{\PAR_1}, \ldots, L_{\PAR_{\ell_2}}$ by the list of $\TENS$-links 
$L'_{\TENS_1}, \ldots, L'_{\TENS_{\ell_2}}$ simultaneously. 
\end{definition}
It is obvious that $\Theta$ and $\tpex(\Theta, L_{\TENS}, L_{\PAR})$ belong to the same PS-family. 
Moreover,\\
 $\tpex(\tpex(\Theta, L_{\TENS}, L_{\PAR}), L'_{\TENS}, L'_{\PAR})$ is $\Theta$. 
Then for each two proof structures $\Theta_1$ and $\Theta_2$, we define a relation 
$\Theta_1 \Leftrightarrow \Theta_2$ if 
there are $\TENS$-link $L_{\TENS}$ and $\PAR$-link $L_{\PAR}$ in $\Theta_1$ such that
$\Theta_2$ is $\tpex(\Theta_1, L_{\TENS}, L_{\PAR})$. 
Then $\Leftrightarrow$ is a symmetric relation from the observation above. 
On the other hand, if $\Theta$ is an MLL proof net and $\Theta \Leftrightarrow \Theta'$, 
then $\Theta'$  is not always an MLL proof net. 
Figure~\ref{figTPExchangeCounterExample} shows such an example. 
Theorem~\ref{mainTheorem1} below describes a necessary and sufficient condition that $\Theta'$ is an MLL proof net. \\
As to general $\TENS$-$\PAR$-exchange $\tpex(\Theta, \langle L_{\TENS_1}, \ldots, L_{\TENS_{\ell_1}} \rangle, \langle L_{\PAR_1}, \ldots, L_{\PAR_{\ell_2}} \rangle)$, 
note that we do not assume that 
each element of 
$\langle L_{\TENS_1}, \ldots, L_{\TENS_{\ell_1}} \rangle$ 
(resp. $\langle L_{\PAR_1}, \ldots, L_{\PAR_{\ell_2}} \rangle$)
does not appear in $\Theta$ like substitution of $\lambda$-calculus, because of convenience. 
In addition, note that Proposition~\ref{propMainKeyObserve} states
when $\Theta$ is an MLL proof net and 
$L_{\TENS} : \frac{A \quad B}
{A \TENS B}$  
(resp. $L_{\PAR} : \frac{C \quad D}
{C \PAR D}$ ) appears in $\Theta$, then 
$\tpex(\Theta, \langle L_{\TENS} \rangle , \langle \rangle)$ 
(resp. $\tpex(\Theta, \langle \rangle , \langle L_{\PAR} \rangle)$) is not an MLL proof net
(although these two belong to the same PS-family as $\Theta$). 
\\
Moreover 
from Corollary~\ref{corPS-familyPNsInvariant}, we can easily see that 
if $\Theta_1$ and $\Theta_2$ are MLL proof nets that belong to the same PS-family, then there
is a sequence of proof structures $\Theta'_1, \ldots, \Theta'_k \, (k \ge 0)$
such that $\Theta_1 \Leftrightarrow \Theta'_1 \Leftrightarrow \cdots \Leftrightarrow \Theta'_k \Leftrightarrow \Theta_2$.
Theorem~\ref{mainTheorem2} below says that 
we can always find such a sequence $\Theta'_1, \ldots, \Theta'_k$ such that each element $\Theta'_i \, (1 \le i \le k)$ 
is an {\bf MLL proof net}. This does not seem trivial.

\subsection{Main Theorems}
In this section, we answer the following question: ``in our framework is error-correcting possible?'' 
Our answer is negative. Corollary~\ref{corImpossibilityOfErrorCorrecting}
says that this is impossible even for 
one error-correcting. \\
Before that, we state a characterization of the condition $d_{\mathcal F}(\Theta_1, \Theta_2)=2$,
where ${\mathcal F}$ is a PS-family and  $\Theta_1$ and $\Theta_2$ are MLL proof nets 
belonging to ${\mathcal F}$. The characterization is used in the proof of Lemma~\ref{lemmaMain}
of Appendix~\ref{secLemmaMain}, which is needed to prove Theorem~\ref{mainTheorem2}. 
\begin{theorem}
\label{mainTheorem1}
Let $\Theta$ be an MLL proof net.
Moreover let $L_{1\TENS}: 
          \frac{A \quad B}{A \TENS B}$ 
 and $L_{2\PAR}: 
          \frac{C \quad D}{C \PAR D}$ be a $\TENS$-link and a $\PAR$-link in $\Theta$ respectively. 
Then $\tpex(\Theta, L_{1 \TENS}, L_{\PAR2})$ is an MLL proof net iff one of the followings holds in $\Theta$:
\begin{enumerate}
\item [(1)] $C$ is a conclusion of $e_\Theta(A)$ and 
$D$ is a conclusion of $e_\Theta(B)$;
\item [(2)] $D$ is a conclusion of $e_\Theta(A)$ and 
$C$ is a conclusion of $e_\Theta(B)$.
\end{enumerate}
\end{theorem}

Our proof of Theorem~\ref{mainTheorem1} is given in Appendix~\ref{secProofOfMainTheorem1}. 

\begin{theorem}
\label{mainTheorem2}
Let $\Theta$ and $\Theta'$ be two MLL proof nets belonging to the same PS-family ${\mathcal F}$. 
Then there is $n \in \mathbb{N}$ and a sequence of MLL proof nets $\Theta_1, \ldots, \Theta_n$ such that
\[ \Theta \Leftrightarrow \Theta_1 \Leftrightarrow \cdots \Leftrightarrow \Theta_n \Leftrightarrow \Theta'.\]
\end{theorem}

\begin{proof}
We assume that $\Theta$ and $\Theta'$ are MLL proof nets, 
but we do not have such a sequence of MLL proof nets for any $n \in \mathbb{N}$. 
Moreover we can choose two MLL proof nets $\Theta$ and $\Theta'$ in ${\mathcal F}$
such that 
there is no MLL proof net $\Theta''$ such that 
$d(\Theta, \Theta'') < d(\Theta, \Theta')$ and $d(\Theta'', \Theta') < d(\Theta, \Theta')$ 
since it is sufficient to prove the theorem. 
Then from Corollary~\ref{corPS-familyPNsInvariant}, we can easily deduce that $d_{\mathcal F}(\Theta, \Theta')$ is even, i.e.,  $d_{\mathcal F}(\Theta, \Theta') = 2m$. 
In addition there are 
$m \, \, \TENS$-links $L_{\TENS 1}:\frac{A_1 \quad B_1}{A_1 \TENS B_1}, \ldots, L_{\TENS m}:\frac{A_m \quad B_m}{A_m \TENS B_m}$ in $\Theta$ and
$m \, \, \PAR$-links $L_{\PAR 1}: \frac{C_1 \quad D_1}{C_1 \PAR D_1}, \ldots, L_{\PAR m}:\frac{C_m \quad D_m}{C_m \PAR D_m}$ in $\Theta$ such that 
$\Theta'$ is $\tpex(\Theta, \langle L_{\TENS 1}, \ldots, L_{\TENS m} \rangle , 
                            \langle L_{\PAR 1}, \ldots, L_{\PAR m} \rangle )$.
Let $\Theta_{i,j} \, (1 \le i, j \le m)$  be $\tpex(\Theta, L_{\TENS i}, L_{\PAR j})$. 
Then our assumption means that $\Theta_{i,j}$ is not an MLL proof net for any $i,j \, (1 \le i, j \le m)$
(The assumption is used in the proof of Lemma~\ref{lemmaMain} of Appendix~\ref{secLemmaMain}). 
Then we derive a contradiction from these settings by induction on 
lexicographic order $\langle m, |{\mathbb{L}}_{\Theta}| \rangle$, where 
$|{\mathbb{L}}_{\Theta}|$ is the number of link occurrences in $\Theta$. 
\begin{enumerate}
\item [(1)] The case where $m=0$ and $m=1$: \\
  It is obvious.
\item [(2)] The case were $m>1$:
\begin{enumerate}
\item [(a)] The case where $\Theta$ consists of exactly one ID-link:\\
  In this case there is neither a $\TENS$-link nor a $\PAR$-link in $\Theta$. This is a contradiction to $m>1$. 
\item [(b)] The case where $\Theta$ includes a $\PAR$-formula $C \PAR D$ as a conclusion:\\
  We choose such a $\PAR$-link $L_{\PAR}: \frac{C \quad D}{C \TENS D}$. 
  \begin{enumerate}
  \item [(i)] The case where $C \PAR D$ is not $C_j \PAR D_j$ for any $j \, (1 \le j \le m)$:\\
    Let $\Theta_0$ be $\Theta$ except that $L_{\PAR}$ is eliminated.
    Then we can apply inductive hypothesis to $\Theta_0$ and a subproof net of $\Theta'$, 
    $\tpex(\Theta_0, \langle L_{\TENS 1}, \ldots, L_{\TENS m} \rangle , 
                            \langle L_{\PAR 1}, \ldots, L_{\PAR m} \rangle )$. 
    We derive a contradiction.
  \item [(ii)] The case where $C \PAR D$ is $C_{j_{0}} \PAR D_{j_{0}}$ for some $j_{0} \, (1 \le j_{0} \le m)$:\\
    In this case, by Lemma~\ref{lemmaMain}, $\Theta'$ is not an MLL proof net. This is a contradiction.
  \end{enumerate}
\item [(c)] The case where the conclusions of $\Theta$ do not have any $\PAR$-formula:\\
  In this case, by Splitting lemma (Lemma~\ref{lemmaSplitting}), 
  we have a $\TENS$-conclusion $A \TENS B$ and its $\TENS$-link $L_{A \TENS B}$ in $\Theta$ such that 
  $\Theta$ is decomposed into $e_{\Theta}^{PN}(A)$, $e_{\Theta}^{PN}(B)$, and $\TENS$-link $L_{A \TENS B}$ 
  \begin{enumerate} 
  \item [(i)] The case where $A \TENS B$ is not $A_i \TENS B_i$ for any $i \, (1 \le i \le m)$:\\
  In this case if the number of $\PAR$-links from $L_{\PAR 1}, \ldots, L_{\PAR m}$ in $e_{\Theta}(A)$ is the same 
  as the number of $\TENS$-links from $L_{\TENS 1}, \ldots, L_{\TENS m}$ in $e_{\Theta}(A)$, then 
  we can apply inductive hypothesis to $e_{\Theta}(A)$ and a subproof net of $\Theta'$, 
  $\tpex(e_{\Theta}(A), \langle L_{\TENS 1}, \ldots, L_{\TENS m} \rangle , 
                            \langle L_{\PAR 1}, \ldots, L_{\PAR m} \rangle)$. 
  Then we derive a contradiction.
  Otherwise, let $\Theta'_A$ be   $\tpex(e_{\Theta}(A), \langle L_{\TENS 1}, \ldots, L_{\TENS m} \rangle , 
                            \langle L_{\PAR 1}, \ldots, L_{\PAR m} \rangle)$. 
  Then by Corollary~\ref{corPS-familyPNsInvariant}, $\Theta'_A$ is not an MLL proof net. 
  Therefore $\Theta'$ is not an MLL proof net. This is a contradiction. 
  \item [(ii)] The case where $A \TENS B$ is $A_i \TENS B_i$ for some $i \, (1 \le i \le m)$:\\    
  Then we can find a DR-switching $S'$ for $\Theta'$ such that $S'(\Theta')$ is disconnected since 
  The $\TENS$-link $L_{\TENS i}$ is replaced by a $\PAR$-link $L_{\TENS i}$.  
  Therefore $\Theta'$ is not an MLL proof net. This is a contradiction. 
  \end{enumerate} 
\end{enumerate}
\end{enumerate}
Therefore, for some $i_0, j_0 \, (1 \le i_0, j_0 \le m)$, 
$\Theta_{i_0, j_0} (= \tpex(\Theta, L_{\TENS i_0}, L_{\PAR j_0}))$ is an MLL proof net. We have done. 
$\Box$
\end{proof}

\begin{lemma}
\label{lemmaMain}
The assumptions are inherited from the case (2-b-ii) of the proof above of Theorem~\ref{mainTheorem2}. 
Then, 
$\Theta'=\tpex(\Theta, \langle L_{\TENS 1}, \ldots, L_{\TENS m} \rangle , 
               \langle L_{\PAR 1}, \ldots, L_{\PAR j_0}, \ldots, L_{\PAR m} \rangle )$ is not an MLL proof net.
\end{lemma}

A proof of the lemma is given in Appendix~\ref{secLemmaMain}.\\
When a PS-family ${\mathcal F}$ has at least two MLL proof nets, 
we define the distance $d({\mathcal F})$ of ${\mathcal F}$ itself in the usual manner:
\[ d({\mathcal F}) = \min \{ d_{{\mathcal F}}(\Theta_1, \Theta_2) \, | \, \Theta_1, \Theta_2 \in {\mathcal F} 
\wedge (\Theta_1 \, \mbox{and} \, \Theta_2 \, \mbox{are MLL proof nets}) \wedge \Theta_1 \neq \Theta_2 \} \] 
Then from Theorem~\ref{mainTheorem2} the following corollary is easily derived.

\begin{corollary}
\label{corImpossibilityOfErrorCorrecting}
For any PS-family ${\mathcal F}$, if the number of the MLL proof nets in ${\mathcal F}$ is equal to or greater than 2, 
then $d({\mathcal F}) = 2$. 
\end{corollary}
Corollary~\ref{corImpossibilityOfErrorCorrecting} means that 
one error-correcting is impossible for any PS-family of MLL. 

\begin{figure}[htbp]
\begin{center}
\includegraphics[scale=.5]{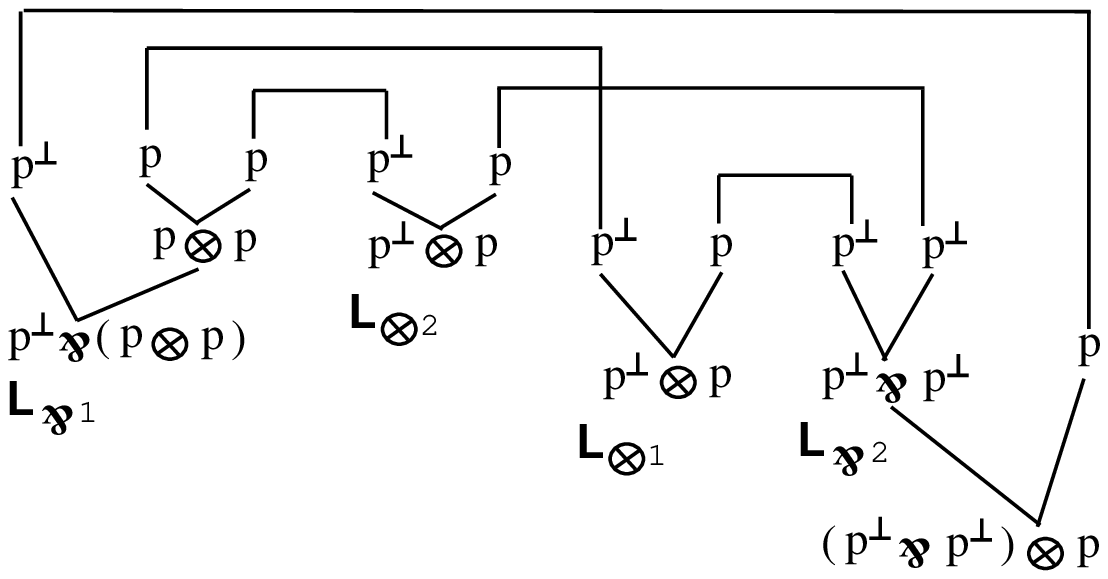}
\quad
\includegraphics[scale=.5]{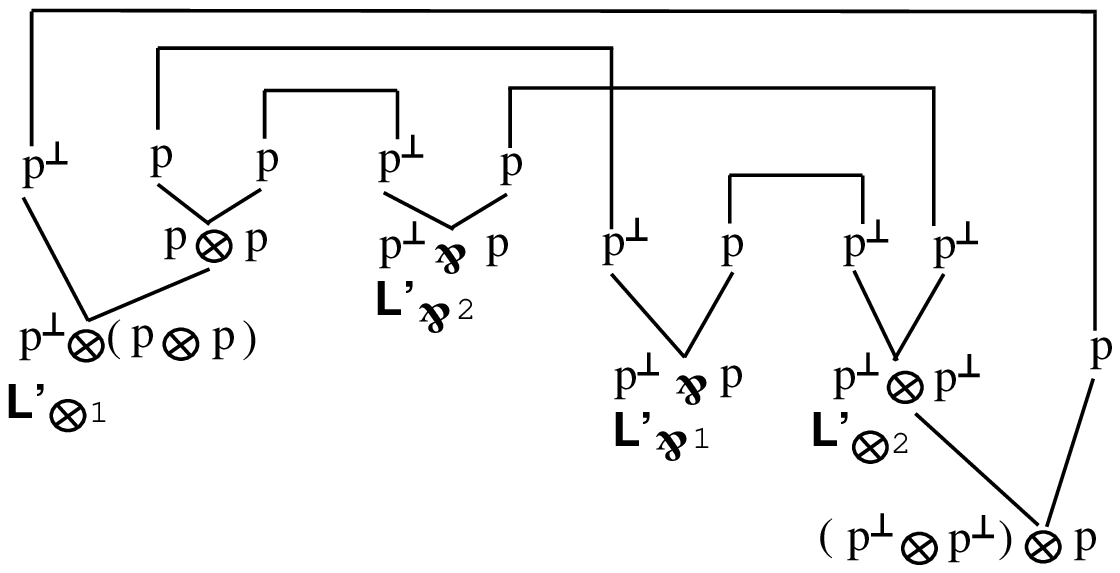}
\caption[MLL proof nets $\Theta$ and $\Theta'$]{MLL proof nets $\Theta$ and  $\Theta'$}  
\label{ex1}
\end{center}
\end{figure}

\begin{figure}[htbp]
\begin{center}
\includegraphics[scale=.5]{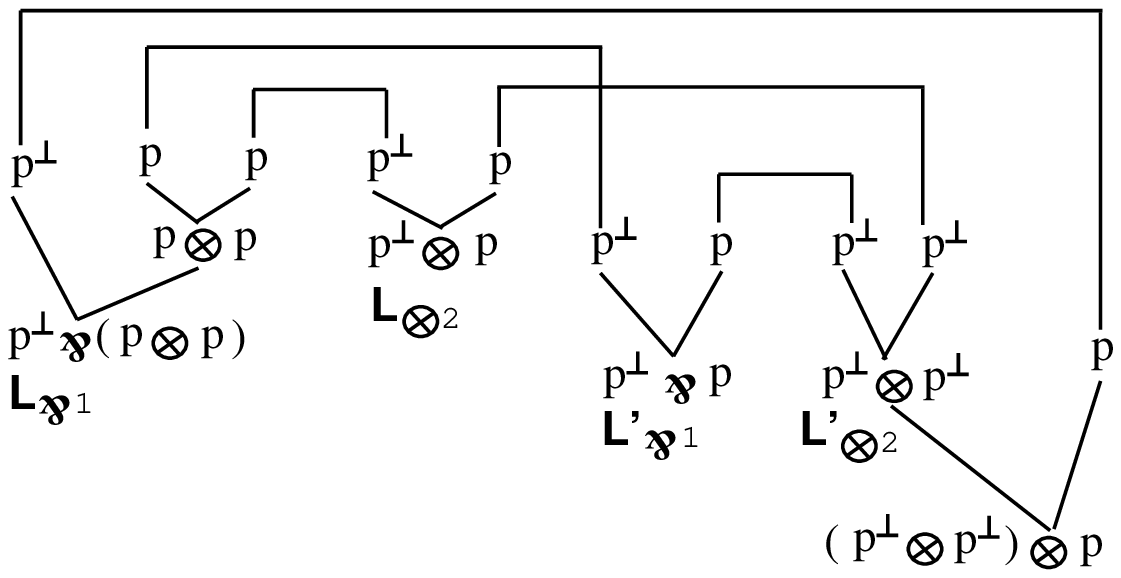}
\quad
\includegraphics[scale=.5]{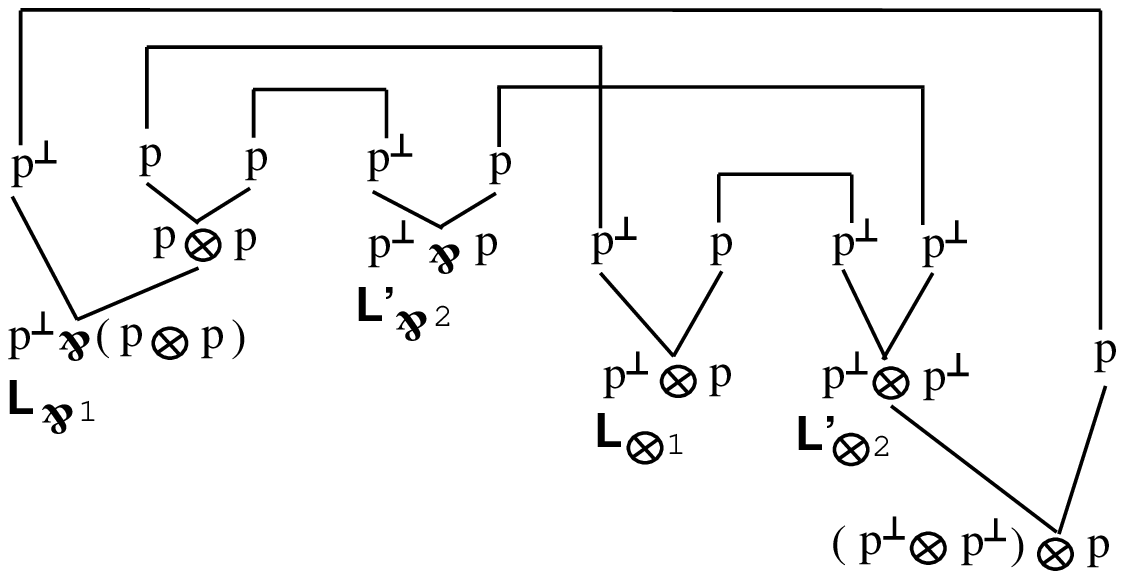}
\caption[MLL proof nets $\Theta_1$ and $\Theta_2$]{MLL proof nets $\Theta_1$ and $\Theta_2$}  
\label{ex2}
\end{center}
\end{figure}

\begin{example}
\label{exMainEx1}
Our proof of Theorem~\ref{mainTheorem2} states that when $\Theta$ and $\Theta'$ are MLL proof nets 
belonging to the same PS-family ${\mathcal F}$ and $d_{\mathcal F}(\Theta, \Theta') \ge 2$, 
we can always find an MLL proof net $\Theta''$ such that 
$d_{\mathcal F}(\Theta, \Theta'')=2$ and $d_{\mathcal F}(\Theta'', \Theta')=d_{\mathcal F}(\Theta, \Theta')-2$. 
We show an example in the following. \\
For two MLL proof nets $\Theta$ of the left side of Figure~\ref{ex1} and $\Theta'$ of the right side of Figure~\ref{ex1} belonging to
the same PS-family, 
$d(\Theta, \Theta')=4$ holds. 
Then when we let the left side of Figure~\ref{ex2} be $\Theta_1$, then
$\Theta_1 = \tpex(\Theta, L_{\TENS 1}, L_{\PAR 2})$ (and $\Theta = \tpex(\Theta_1, L'_{\TENS 2}, L'_{\PAR 1})$). 
Moreover we find $d(\Theta, \Theta_1)=2$ and $d(\Theta_1, \Theta')=2$.
But such a $\Theta_1$ is not unique. 
In fact when we let $\Theta_2$ be the right side of Figure~\ref{ex2},
then $\Theta_2 = \tpex(\Theta, L_{\TENS 2}, L_{\PAR 2})$ (and $\Theta = \tpex(\Theta_2, L'_{\TENS 2}, L'_{\PAR 2})$). 
By the way, the PS-family has nine MLL proof nets. \\
{\bf Warning:} This example is not a substitute for Corollary~\ref{corImpossibilityOfErrorCorrecting}. 
The statement of Corollary~\ref{corImpossibilityOfErrorCorrecting} is a {\it universal} one. 
Therefore one example is not enough to prove the statement. 
\end{example}

\subsection{Other Topics}
\label{subsecOtherTopics}
In this section we discuss ongoing research directions in our framework.
\subsubsection{The Number of MLL Proof Nets in a PS-family}
It is interesting to consider how many MLL proof nets a given PS-family has. 
We have a characterization of the PS-families without any MLL proof nets as 
an elementary result. \\
Firstly we note that the number of the multiplicative links in an element of a given PS family ${\mathcal F}$ is 
always the same. 

\begin{definition}[PS-connected]
\label{defPSConnected}
Let ${\mathcal F}$ be a PS-family. 
Then ${\mathcal F}$ has the element $\Theta_{\TENS}$ that has only $\TENS$-links as its multiplicative links
(if any). Then there is exactly one DR-switching $S$ for $\Theta_{\TENS}$ that is 
empty set. 
${\mathcal F}$ is PS-connected if 
the unique DR-graph $S(\Theta_{\TENS})$ is connected.
\end{definition}

\begin{proposition}
\label{propCharaMLLPNsZero}
Let ${\mathcal F}$ be a PS-family.
Then ${\mathcal F}$ does not have any MLL proof nets iff 
${\mathcal F}$ is not PS-connected. 
\end{proposition}

\begin{proof}
\begin{enumerate}
\item If part:\\
We assume that that ${\mathcal F}$ is not PS-connected. 
We can easily see that for each element $\Theta$ of ${\mathcal F}$ and 
each DR-switching $S$ for $\Theta$, the DR-graph $\Theta_{S}$ is disconnected. 
Therefore, there is no MLL proof nets in ${\mathcal F}$. 
\item Only-if part:\\
We prove that if ${\mathcal F}$ is PS-connected, then 
${\mathcal F}$ has at least one MLL proof nets 
by induction on the number $n$ of the multiplicative links in ${\mathcal F}$. 
\begin{enumerate}
\item The case where $n=0$:\\
${\mathcal F}$ is PS-connected, ${\mathcal F}$ must be the singleton consisting of exactly one ID-link. 
Therefore ${\mathcal F}$ has exactly one MLL proof net.
\item The case where $n>0$:
\begin{enumerate}
\item The case where there is an element $\Theta$ of ${\mathcal F}$ such that 
by removing one multiplicative link $L: \frac{A \quad B}{A @ B}$ of $\Theta$ and its conclusion $A @ B$, 
two disjoint proof structures $\Theta_1$ with a conclusion $A$ and $\Theta_2$ with a conclusion $B$ is obtained:\\
Let ${\mathcal F_1}$ and ${\mathcal F_2}$ be the PS-families that $\Theta_1$ and $\Theta_2$ belong to respectively.
Both ${\mathcal F_1}$ and ${\mathcal F_2}$ are PS-connected. 
Therefore by inductive hypothesis ${\mathcal F_1}$ and ${\mathcal F_2}$ have MLL proof nets $\Theta'_1$ and $\Theta'_2$ 
respectively. Then let $\Theta'$ be the proof structure obtained from $\Theta'_1$ and $\Theta'_2$ 
by connecting $A$ and $B$ via $\TENS$-link $L': \frac{A \quad B}{A \TENS B}$. 
Then it is obvious that $\Theta'$ is an MLL proof net and $\Theta'$ is an element of ${\mathcal F}$. 
\item Otherwise:\\
Then there is an element $\Theta$ of ${\mathcal F}$ such that 
by removing one multiplicative link $L: \frac{A \quad B}{A @ B}$ of $\Theta$ and its conclusion $A @ B$, 
one proof structure $\Theta_0$ with conclusions $A$ and $B$ is obtained. 
Let ${\mathcal F_0}$ be the PS-family that $\Theta_0$ belongs to. 
${\mathcal F_0}$ is PS-connected. 
Therefore by inductive hypothesis ${\mathcal F_0}$ has an MLL proof net $\Theta'_0$. 
Then let $\Theta'$ be the proof structure obtained from $\Theta'_0$
by connecting $A$ and $B$ via $\PAR$-link $L': \frac{A \quad B}{A \PAR B}$. 
Then it is obvious that $\Theta'$ is an MLL proof net and $\Theta'$ is an element of ${\mathcal F}$. 
$\Box$
\end{enumerate}
\end{enumerate}
\end{enumerate}
\end{proof}

But it is not so easy to give a similar characterization of PS-families with exactly $m$ MLL proof nets 
for a given $m \, (\ge 1)$. At this moment we just obtain the following elementary result.

\begin{proposition}
\label{propInfinitePSFamiliesMLLPNsN}
For any positive integer $m$, there are denumerable PS-families with exactly $m$ MLL proof nets.
\end{proposition}

\begin{figure}[htbp]
\begin{center}
\includegraphics[scale=.5]{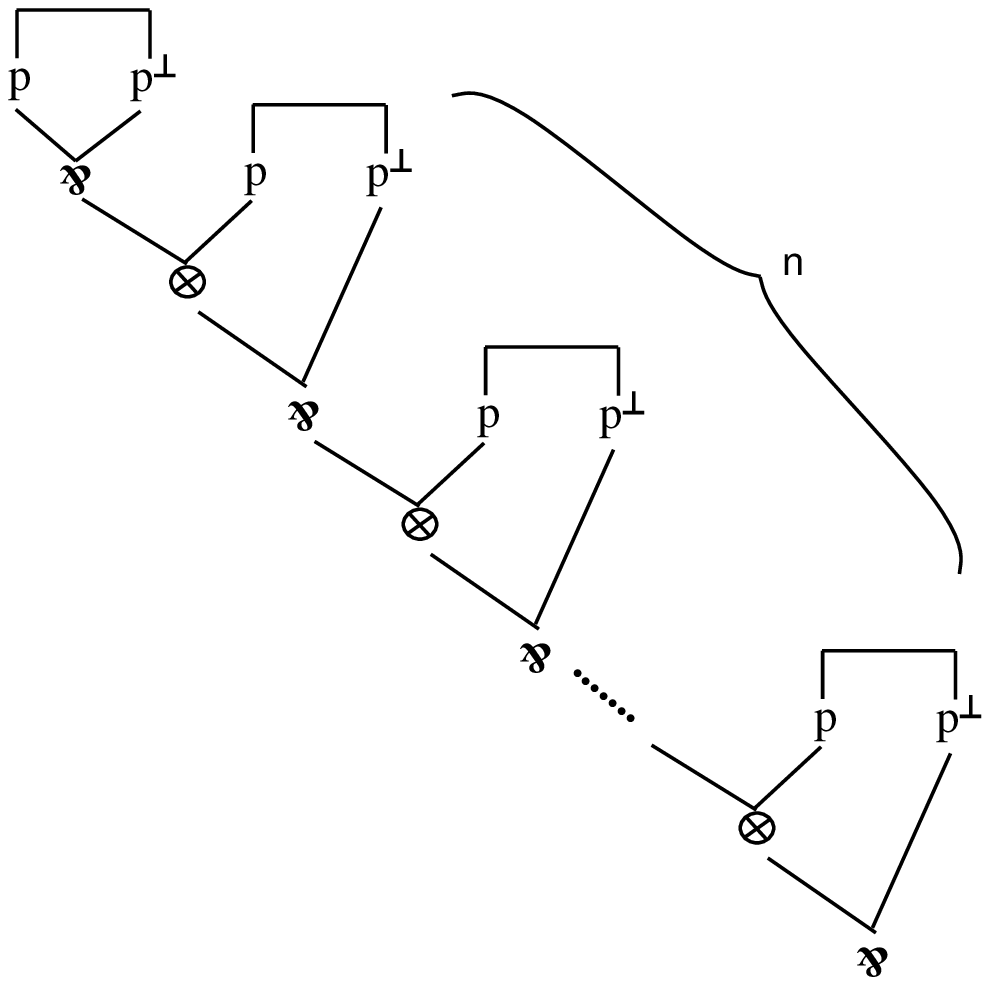}
\quad
\includegraphics[scale=.5]{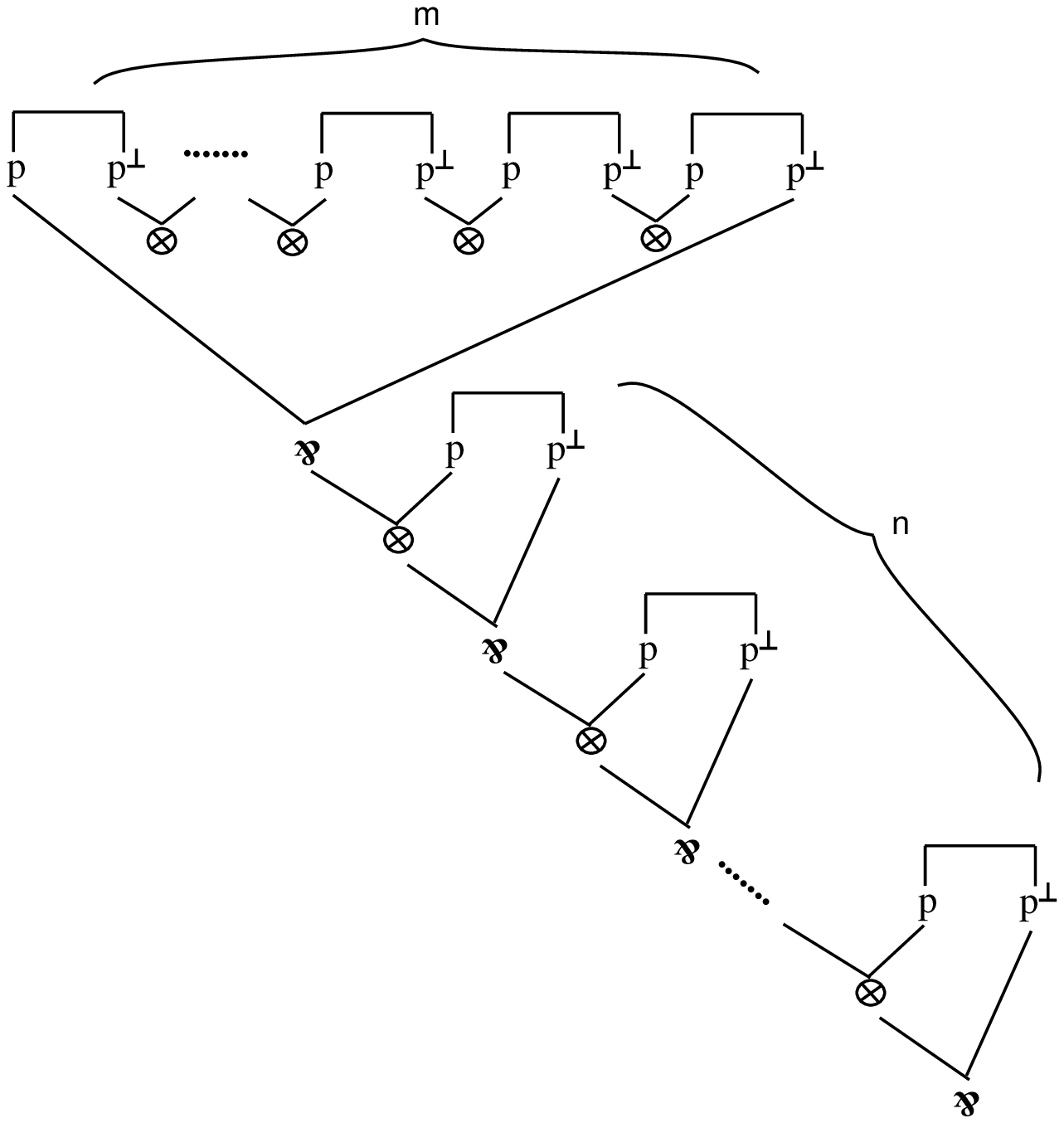}
\caption[Witnesses for Proposition~\ref{propInfinitePSFamiliesMLLPNsN}]{Witnesses for Proposition~\ref{propInfinitePSFamiliesMLLPNsN}}
\label{denumPNs}
\end{center}
\end{figure}

\begin{proof}
If $m=1$, then it is enough to see the left side of Figure~\ref{denumPNs} in order to confirm that the statement is correct. 
Similarly if $m>1$, it is enough to see the right side of Figure~\ref{denumPNs} for the same purpose. 
$\Box$
\end{proof}
But it seems difficult to obtain a characterization of the PS-families even with exactly one MLL proof net. 
The reason is as follows:
\begin{enumerate}
\item There are primitive patterns of such PS-families. 
\item Moreover by combining such primitive patterns appropriately, we can get compound PS-families with 
      exactly one MLL proof net.
\end{enumerate}
In order to get such a characterization, it seems that an appropriate language that describes (denumerable) sets 
of PS-families is needed like the regular language for describing sets of words. 
But since the purpose of this paper is to introduce 
the new notion of PS-families and metric spaces associated with them, 
the question is left open as an interesting one. 

\subsubsection{The Composition of PS-families}
MLL proof nets are composable: we get a MLL proof net by connecting two MLL proof nets via Cut-link.
But this is not the case about MLL proof structures: we may obtain a vicious circle 
by connecting two MLL proof structures via Cut-link (see Section 11.2.6 of \cite{Gir06}).
Therefore we need a care about the composition of PS-families because 
a PS-family always includes MLL proof structures that are not MLL proof nets. 
Moreover this issue is closely related to recent works of Samson Abramsky and his colleagues 
about compact closed categories (For example, see \cite{Abr07}). 
But since the paper is already long, the issue will be treated elsewhere.

\section{Concluding Remarks}
In this paper, we introduced the notion of PS-families over MLL proof structures and 
metric spaces with associated with them. Moreover we proved that in the case where 
A PS-family has more than two MLL proof nets, the distance of the PS-family is 2. \\
Although our main result is the impossibility of one error-correcting in MLL, 
the remedy is possible. By introducing general $\TENS_n$-links and $\PAR_n$-links \cite{DR89}, where 
$n \ge 3$ and these general links have $n$ premises instead of exactly two premises, 
we can construct a PS-family ${\mathcal F}$ such that $d({\mathcal F}) = n$. 
For example, when let $\Theta_1$ (resp. $\Theta_2$ be the general MLL proof net of the left (resp right) side, 
$d_{\mathcal F}(\Theta_1, \Theta_2) = 4$, where ${\mathcal F}$ is the PS-family belonging to $\Theta_1$ and $\Theta_2$. 
Moreover it is obvious that $d({\mathcal F})=4$. 
But at this moment we are not sure whether such an easy modification makes good codes 
(although our main purpose is not to find good codes from PS-families). 
Nevertheless, we believe that Theorem~\ref{mainTheorem2} is a  fundamental theorem in this direction of study, 
because a general version of Theorem~\ref{mainTheorem2} seems to be derived in the extended framework. 
\begin{figure}[htbp]
\begin{center}
\includegraphics[scale=.5]{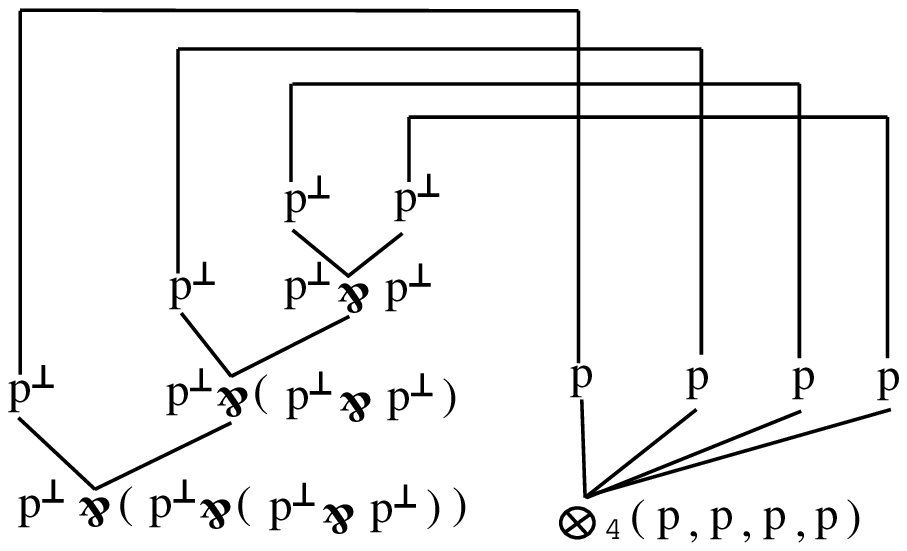}
\quad
\includegraphics[scale=.5]{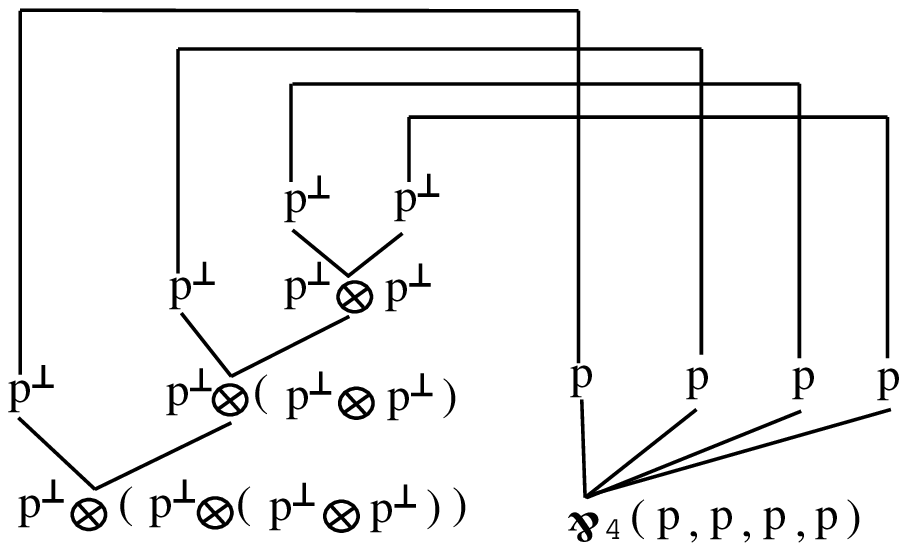}
\caption[General MLL proof nets $\Theta_1$ and $\Theta_2$]{General MLL proof nets $\Theta_1$ and $\Theta_2$}  
\label{general}
\end{center}
\end{figure}

\begin{ack}
The author thanks the participants of the 205th Computer Language Colloquium 
at the Senri office of the Research Center for Semantics and Verification, 
which is a part of AIST. 
He also thanks Lorenzo Tortora de Falco for helpful comments. 
\end{ack}

\appendix
\section{Codes over Binary Words}
In this appendix, we present basic knowledge about codes over binary finite words.
The contents are elementary. The reader can find these materials in any coding theory's textbooks, 
for example \cite{Bay98,MS93}.
The purpose of the appendix is to help the reader understand this paper easily 
by comparing with the standard theory. 
If the reader knows these things already, please ignore the appendix. 

\begin{definition}[Binary Finite Words]
\label{defBinFinWords}
A binary word $w$ with length $n \, (\in \mathbb{N})$ is an element of ${\{ 0, 1 \}}^n$. 
For each $i \, (1 \le i \le n)$, $w[i]$ ($\in \{ 0, 1 \}$) denotes $i$-th element of $w$. 
\end{definition}

\begin{definition}[Distance of Binary Words with the Same Length]
\label{defDisBinFinWords}
Let $w_1$ and $w_2$ be binary words with the same length $n$. 
The the distance of $w_1$ and $w_2$, $d(w_1, w_2)$ is defined as follows:
\[ d(w_1, w_2) = | \{ w_1[i] \in \{ 0, 1 \} \, | \, 1 \le i \le n \wedge w_1[i] \neq w_2[i] \} | \]
\end{definition}
For example, $d(00110, 10011)=3$.

\begin{definition}[Code over Words with Length $n$]
\label{defCodeOverWords}
A code $C$ over words with length $n$ is a subset of ${\{ 0, 1 \}}^n$. 
An element of $C$ is called {\bf codeword}. 
The distance of $C$ is defined as follows:
\[ d(C) = \min \{ d(w_1, w_2) \, | \, w_1, w_2 \in C \wedge w_1 \neq w_2 \} \]
\end{definition}

\begin{example}[Hamming $\langle 7, 4 \rangle$ code]
\label{exHamSevenFour}
The Hamming $\langle 7, 4 \rangle$ code $C$ is the subset of ${\{ 0, 1 \}}^{7}$ such that 
$w \in {\{ 0, 1 \}}^{7}$ is an element of $C$ iff 
$w$ satisfies the following three equations(where $\oplus$ denotes 'exclusive or'):
\begin{eqnarray*}
w[1] \oplus w[2] \oplus w[4] \oplus w[5] = 0 \\
w[2] \oplus w[3] \oplus w[4] \oplus w[6] = 0 \\
w[1] \oplus w[3] \oplus w[4] \oplus w[7] = 0 
\end{eqnarray*}
Then we can easily see $|C|=16$ and $d(C) = 3$ by easy calculation.
As a result the Hamming $\langle 7, 4 \rangle$ code is {\bf one error-correcting}
because when a given $w \in {\{ 0, 1 \}}^{7}$, if $d(w, w')=1$ for some $w' \in C$, then 
for any $w'' (\neq w') \in C$, $d(w, w'')>1$. 
Therefore we can judge that $w$ is $w'$ with one error. 
Moreover the Hamming $\langle 7, 4 \rangle$ code is {\bf two error-detecting}
because when a given $w \in {\{ 0, 1 \}}^{7}$, if $d(w, w')=2$ for some $w' \in C$, then 
for any $w'' (\neq w') \in C$, $d(w, w'') \ge 2$. 
Therefore we can judge that $w$ has exactly two errors. 
But since there may be a different codeword $w''' \in C$ from $w'$ such that $d(w, w''')=2$, 
we can not judge that $w$ is $w'$ with two errors. \\
On the other hand, in the Hamming $\langle 7, 4 \rangle$ code $C$, 
we can not do one error-correcting and two error-detecting at the same time, 
because there are $w_1, w_2 \in C$ and $w' \in {\{ 0, 1 \}}^{7} - C$ such that 
$d(w_1, w')=2$ and $d(w', w_2)=1$. 
Therefore we can not decide whether $w'$ is $w_1$ with two errors or $w_2$ with one error. 
We have to decide whether we adopt the one error-correcting interpretation or 
the two error-detecting interpretation. 
If we adopt the one error-correcting interpretation, then $w'$ is $w_2$ with one error.
If we adopt the two error-detecting interpretation, then $w'$ has two errors, but 
we can not say $w'$ is $w_1$ with two errors by the reason of the paragraph above. 
\end{example}

\label{secBasicCodeTheory}
\section{Basic Properties of Empires}
\label{secEmpBasicProp}
In this section we prove basic properties of empires. 
These properties are well-known in the literature, for example ~\cite{Gir87,BW95,Gir96,Gir06}. 
\label{subsecAlreadyKnownResults}
Before presenting results, 
we fix terminology about paths of indexed formulas in a DR-graph.

\begin{definition}
\label{defPathImmedAboveAdjacentBelow}
Let $\Theta$ be an MLL proof net, $S$ be a DR-switching for $\Theta$, and 
$A, B \in \fml(\Theta)$.
Then there is a unique path $\theta$ from $A$ to $B$ 
in $\Theta_{S}$. 
We say that $\theta$ {\bf passes immediately above or adjacent to} $A$ (resp. $B$)
if
$\theta$ includes a formula $C$ such that 
there is the link $L$ whose conclusion is $A$ (resp. $B$) and 
$C$ is a premise or another conclusion of $L$.
We say that $\theta$ {\bf passes immediately below} $A$ (resp. $B$) if 
$\theta$ includes a formula $C$ such that
there is the link $L$ whose premise is $A$ (resp. $B$) and 
$C$ is the conclusion of $L$.
\end{definition}

\begin{proposition}
\label{propEmpireIncudinglPremises}
Let $B \in e_\Theta(A)$ and 
$L \in {\mathbb L}_{e_\Theta(A)}$ such that the conclusion of $L$ is $B$. 
Then if $B'$ is a premise or a conclusion of $L$, then 
$B' \in e_\Theta(A)$. 
\end{proposition}

\begin{proof}
We prove this by case analysis.
If $B'=B$, then it is obvious. 
So we assume $B' \neq B$ in the following.  
\begin{enumerate}
\item The case where $L$ is an ID-link:\\
Then $B$ and $B'$ are literals which are dual each other. 
Since $B \in e_\Theta(A)$, 
for each DR-switching $S$, $B \in \fml({\Theta_S}^{A})$. 
Then it is obvious that $B' \in \fml({\Theta_S}^{A})$. 
So, $B' \in e_\Theta(A)$.
\item The case where $L$ is a $\TENS$-link:\\
  Then $B'$ is a premise of $L$. 
  The rest of the proof of this case is similar to the case above. 
\item The case where $L$ is a $\PAR$-link:\\
  Then $B'$ is a premise of $L$. 
  Without loss of generality, we can assume that $B'$ is the left premise of $L$. 
  We assume $B' \not\in e_\Theta(A)$. 
  Then there a DR-switching $S$ such that $B' \not\in \fml({\Theta_S}^{A})$. 
  By the assumption $S$ selects the right premise $B''$ in $L$. 
  Since $\Theta_S$ is acyclic and connected, there is a unique path $\theta$ from $B$
  to $B'$ in $\Theta_{S}$.
  If $\theta$ does not include $A$, then by the definition of $\fml({\Theta_S}^{A})$ and by $B \in \fml({\Theta_S}^{A})$, 
  we derive $B' \in \fml({\Theta_S}^{A})$, which is a contradiction. 
  So, $\theta$ includes $A$ and $\theta$ has two subpaths $\theta_1$ from $B$ to $A$ that passes immediately above or adjacent to $A$ and $\theta_2$ from $A$ to $B'$ that passes immediately below $A$.
  Then if $\theta$ includes $B''$, then $\theta_1$ includes $B''$ and letting $S'$ be $S$ except $S'$ selects $B'$, we obtain $B \not\in \fml({\Theta_{S'}}^{A})$ 
  and then $B \not\in e_{\Theta}(A)$, which is a contradiction. 
  Therefore $\theta$ does not include $B''$. 
  Then let $S'$ be the DR-switching such that $S'$ is $S$ except that 
  $S'$ selects the left premise $B'$ in $L$. 
  Then $\Theta_{S'}$ has a cycle. This is a contradiction. 
  $\Box$
\end{enumerate}
\end{proof}
The following corollary is easily derived from the proposition above.
\begin{corollary}
\label{corempirePNisPS}
The pair $\langle e_\Theta(A), {\mathbb L}_{e_\Theta(A)} \rangle$ is 
an MLL proof structure.
\end{corollary}

\begin{proposition}
\label{propParImperialism2}
If $B_1 \in e_\Theta(A)$, 
$B_2 \not\in e_\Theta(A)$, and 
$L$ is a $\PAR$-link such that $B_1$ and $B_2$ are the premise of $L$, 
then 
the conclusion $B$ of $L$ does not belongs to $e_\Theta(A)$.
\end{proposition}

\begin{proof}
We assume that $B \in e_\Theta(A)$. 
Then by Proposition~\ref{propEmpireIncudinglPremises}, 
$B_2 \in e_\Theta(A)$. 
This is a contradiction. 
$\Box$
\end{proof}

\begin{proposition}
\label{propTensImperialism}
If $B \in e_\Theta(A)$ such that $B \neq A$ and 
$L$ is a $\TENS$-link such that $B$ is a premise of $L$, 
then 
the premises and the conclusion of $L$ belong to $e_\Theta(A)$.
\end{proposition}

\begin{proof}
Similar to the case 2 of the proof of Proposition~\ref{propEmpireIncudinglPremises}. 
$\Box$
\end{proof}

\begin{proposition}
\label{propParImperialism}
If $B_1,  B_2 \in e_\Theta(A)$ such that $B_1 \neq B_2, B_1 \neq A, B_2 \neq A$ and 
$L$ is a $\PAR$-link such that $B_1$ and $B_2$ are the premises of $L$, 
then 
the conclusion $B$ of $L$ belongs to $e_\Theta(A)$.
\end{proposition}

\begin{proof}
From the assumption for each DR-switching $S$ for $\Theta$, 
$B_1,  B_2 \in \fml({\Theta_S}^{A})$. 
If $S$ selects $B_1$ in $L$, then 
there is an edge between $B_1$ and $B$ in ${\Theta_S}^{A}$.
That is $B \in \fml({\Theta_S}^{A})$. 
On the other hand, if $S$ selects $B_2$ in $L$, then 
there is an edge between $B_2$ and $B$ in ${\Theta_S}^{A}$.
That is $B \in \fml({\Theta_S}^{A})$. 
Hence $B \in e_\Theta(A)$. 
$\Box$
\end{proof}

Next, we prove that there is a DR-switching $S$ such that 
$\fml({\Theta_S}^{A}) = e_\Theta(A)$. 

\begin{definition}
\label{defPrincipalSwitching}
Let $S$ be a DR-switching for an MLL proof net $\Theta$ including $A$.
we say that  $S$ is a principal DR-switching (or simply principal switching) for $A$ in $\Theta$ 
if $S$ satisfies the following conditions:
\begin{enumerate}
\item if there is a $\PAR$-link $L$ such that a premise of $L$ is $A$, 
      then $S$ selects $A$, not the other premise of $L$ in $L$ and 
\item if there is a $\PAR$-link $L$ such that one premise $B_1$ of $L$  belongs to  $e_\Theta(A)$ and 
the other premise $B_2$ of $L$ does not belong to $e_\Theta(A)$, 
then $S$ selects $B_2$ in $L$. 
\end{enumerate}
\end{definition}

When a given MLL proof net $\Theta$ and a formula $A$ in $\Theta$, 
we can easily see that we can always find a principal DR-switching for $A$ in $\Theta$ from the definition above,
since if we find a $\PAR$-link satisfying any of the assumptions of the conditions, then 
we can always choose the switch for the $\PAR$-link that satisfies the conditions. 

\begin{proposition}
\label{propPrincipalSwitching}
Let $S$ be a DR-switching for an MLL proof net $\Theta$. 
Then $S$ is a principal DR-switching for a formula $A$ in $\Theta$ $\, \, $ iff 
$\, \,$ $\fml({\Theta_S}^{A}) = e_\Theta(A)$. 
\end{proposition}

\begin{proof}
The if-part is obvious. Hence we concentrate on the only-if part in the following.\\
Let $S$ be a principal DR-switching. 
It is obvious that $e_\Theta(A) \subseteq \fml({\Theta_S}^{A})$ 
from the definition of empires. 
In order to prove $\fml({\Theta_S}^{A}) \subseteq e_\Theta(A)$, 
we need the following claim.

\begin{claim}
\label{claimPropPrincipalSwitching}
Let $B \in \fml({\Theta_S}^{A})$.
If the unique path $\theta$ from $A$ to $B$ in 
${\Theta_S}^{A}$ includes a $\PAR$-formula $C \PAR D$, then 
$C$ and $D$ must belong to $e_\Theta(A)$.
\end{claim}

{\it Proof of Claim~\ref{claimPropPrincipalSwitching}.} \ \ 
We prove the claim by induction on the number of $\PAR$-formulas in $\theta$.\\
If $\theta$ does not include any $\PAR$-formula, then the claim is obvious. \\
Let $C \PAR D$ be the nearest $\PAR$-formula to $B$ in $\theta$ and
$E$ be the formula immediately before $C \PAR D$ in $\theta$. 
Then we consider the subpath $\theta'$ of $\theta$ from $A$ to $E$.
Then the number of $\PAR$-formulas in $\theta'$ is less than that of $\theta$. 
So by inductive hypothesis, the premises of each $\PAR$-formula in $\theta'$ belong to 
$e_\Theta(A)$. 
Then from Proposition~\ref{propEmpireIncudinglPremises}, Proposition~\ref{propTensImperialism}, and Proposition~\ref{propParImperialism}, the formulas in $\theta'$ must belong to $e_\Theta(A)$.
So $E \in e_\Theta(A)$. 
Then the following two cases are considered:
\begin{enumerate}
\item The case where $E$ is either $C$ or $D$:\\
  Without loss of generality, we can assume that $E$ is $C$. 
  Then we assume that $C \PAR D \not\in e_\Theta(A)$. 
  But this contradicts that $S$ is a principal DR-switching. 
\item The case where $E$ is neither $C$ nor $D$:\\
  Since $E \in e_\Theta(A)$, 
  from Proposition~\ref{propEmpireIncudinglPremises} we can derive 
  $C \PAR D \in e_\Theta(A)$. 
  Then again by Proposition~\ref{propEmpireIncudinglPremises}, 
  $C$ and $D$ must belong to $e_\Theta(A)$. 
\ {\it the end of proof of Claim~\ref{claimPropPrincipalSwitching}.} 
\end{enumerate}
{\it the end of proof of Claim~\ref{claimPropPrincipalSwitching}}\\
Hence using the claim, from Proposition~\ref{propEmpireIncudinglPremises}, Proposition~\ref{propTensImperialism}, and Proposition~\ref{propParImperialism}, we can derive $B \in e_\Theta(A)$. 
$\Box$
\end{proof}

\begin{corollary}
\label{corEmpirePN}
$\langle e_\Theta(A), {\mathbb L}_{e_\Theta(A)} \rangle$ is an 
MLL proof net.
\end{corollary}

\begin{proof}
Since $\Theta' = \langle e_\Theta(A), {\mathbb L}_{e_\Theta(A)} \rangle$ is 
a proof structure by Corollary~\ref{corempirePNisPS}, we concentrate on 
the correctness criterion. 
Let $S'$ be a DR-switching for $\langle e_\Theta(A), {\mathbb L}_{e_\Theta(A)} \rangle$. 
Then there is a principal DR-switching $S$ for $A$ in $\Theta$ which is an extension of $S'$. 
Then by Proposition~\ref{propPrincipalSwitching}, 
$\fml({\Theta_{S}}^{A}) = e_\Theta(A) = \fml({\Theta'}_{S'})$. 
Therefore ${\Theta_{S}}^{A} = {\Theta'}_{S'}$. 
This means that ${\Theta'}_{S'}$ is acyclic and connected. $\Box$
\end{proof}

\begin{corollary}
\label{corEmpireGreatest}
$\langle e_\Theta(A), {\mathbb L}_{e_\Theta(A)} \rangle$ is the 
greatest MLL sub-proof net of $\Theta$ among the MLL sub-proof nets of $\Theta$ with a conclusion $A$.
\end{corollary}

\begin{proof}
Let $\Theta'$ be an MLL sub-proof net of $\Theta$ with conclusion $A$ such that 
$e_\Theta(A) \subsetneq \fml(\Theta')$.
Then, 
if $S$ is a principal switching for $A$ in $\Theta$, then 
by Proposition~\ref{propPrincipalSwitching},
$\fml({\Theta_S}^{A}) = e_\Theta(A)$. So there is a formula $B \in \fml(\Theta')$ such that 
$B \not\in  \fml({\Theta_S}^{A})$. 
Next we consider the MLL proof net $\Theta'$ as the {\it root} proof net instead of $\Theta$. 
Note that for any DR-switching $S'_0$ for $\Theta'$, there is no path $\theta'$ in $\Theta'_{S'_0}$ such that 
$\theta'$ passes immediately below $A$. 
Moreover since $e_\Theta(A) \subsetneq \fml(\Theta')$,  
by extending a principal switching $S_0$ for $A$ in $e_\Theta(A)$, 
we can obtain
a DR-switching $S'_0$ for $\Theta'$. But then $A$ and $B$ are 
disconnected in $\Theta'_{S'_0}$ by the note above. This is a contradiction. 
$\Box$
\end{proof}

\begin{corollary}
\label{corPNConcEmpire}
If $A$ is a conclusion of an MLL proof net $\Theta$, then 
$\langle e_\Theta(A), {\mathbb L}_{e_\Theta(A)} \rangle = \Theta$. 
\end{corollary}

\begin{corollary}
\label{corPNConcEmpire2}
If $B$ is a conclusion of $e_\Theta(A)$, then $A \in e_\Theta(B)$ 
(but $A$ is not necessarily a conclusion of $e_\Theta(B)$). 
\end{corollary}

\begin{proposition}
\label{propEmpireIntersectionEmpty}
If $B \not\in e_\Theta(A)$ and 
$A \not\in e_\Theta(B)$, then 
$e_\Theta(A) \cap e_\Theta(B) = \emptyset$. 
\end{proposition}

\begin{proof}
We derive a contradiction from assumptions
$B \not\in e_\Theta(A)$, 
$e_\Theta(A) \cap e_\Theta(B) \neq \emptyset$, and
$A \not\in e_\Theta(B)$.  
We assume that $C \in e_\Theta(A) \cap e_\Theta(B)$.
We claim the following.
\begin{claim}
\label{claimPropEmpireIntersectionEmpty}
There is a principal switching $S^f_{B}$ 
for $B$ such that 
there is no path from $A$ to $B$ in 
$(\Theta_{S^f_{B}})^{A}$. 
\end{claim}
{\it Proof of Claim~\ref{claimPropEmpireIntersectionEmpty}} \ 
Let $S_{B}$ be a principal switching for $B$. 
Then by Proposition~\ref{propPrincipalSwitching}, 
$\fml({(\Theta_{S_{B}})}^{B}) = e_\Theta(B)$. 
Since $A \not\in e_\Theta(B) = \fml({(\Theta_{S_{B}})}^{B})$, in $\Theta_{S_{B}}$ 
there is a unique path $\theta$ from $A$ to $B$ 
in $\Theta_{S_{B}}$ 
such that
$\theta$ passes immediately below $B$. 
Then if each formula in $\theta$ except $A$ is not included in $(\Theta_{S_{B}})^{A}$, then 
we have done. We just let  $S^f_{B}$ be $S_{B}$. 
Next we assume that $\theta$ includes a formula in $(\Theta_{S_{B}})^{A}$ except $A$. 
Then, since $\Theta_{S_{B}}$ is acyclic and connected and by the definition of $(\Theta_{S_{B}})^{A}$, 
$\theta$ from $A$ to $B$ must be included in $(\Theta_{S_{B}})^{A}$.
On the other hand, since $B \not\in e_\Theta(A)$, 
there is a $\PAR$-link $L:\frac{E \quad F}{E \PAR F}$ such that 
exactly one premise of $L$ (i.e., $E$ or $F$) and $E \PAR F$ are not included 
in $e_\Theta(A)$. Without loss of generality we can assume that 
(i) $E \in e_\Theta(A)$,
(ii) $F \not\in e_\Theta(A)$, and (iii) $\theta$ includes the subpath $E, E \PAR F$ 
by picking up the first $\PAR$-link in $\theta$ among such $\PAR$-links. 
Moreover we can show that such a $\PAR$-link is unique in $\theta$ (otherwise, we have a 
$L_0:\frac{E_0 \quad F_0}{E_0 \PAR F_0}$ in $\Theta$ such that 
(i') $E_0 \in e_\Theta(A)$, (ii')$F_0 \not\in e_\Theta(A)$, and (iii') $\theta$ includes the subpath $E_0 \PAR F_0, E_0$
without loss of generality. Then $S_{B}(\Theta)$ has a cycle because $S_{B}(\Theta)$ has a path from $E$ to $E_0$ 
other than the subpath of $\theta$ from $E$ to $E_0$. This is a contradiction). 
\begin{subclaim}
\label{subClaimPropEmpireIntersectionEmpty}
Let $S'_{B}$ be the DR-switching $S_{B}$ except that 
$S'_{B}$ chooses the other formula, i.e., $F$ in $L$. 
Then, ${S'}_{B}$ is a principal switching for $B$.
\end{subclaim}

{\it Proof of Subclaim~\ref{subClaimPropEmpireIntersectionEmpty}} \ 
We suppose not. 
On the other hand, since $S_{B}$ is principal for $B$,
$\theta$ in $(\Theta_{S_{B}})^{A}$ passes immediately above or adjacent to $A$ and 
immediately below $B$ (Otherwise, $\theta$ passes immediately above or adjacent to $B$. 
This means that $A \in \fml(\Theta_{S_{B}}) = e_\Theta(B)$). 
Since $\theta$ includes $E$ and $E \PAR F$, 
we have $E \not\in e_\Theta(B)$
and 
$E \PAR F \not\in e_\Theta(B)$. 
Therefore we must have $F \in e_\Theta(B)$,
because otherwise (i.e. $F\not\in e_\Theta(B)$), it is obvious 
that $S'_{B}$ is a principal switching for $B$.
Since $F \in e_\Theta(B)$, we have a unique path $\theta'$ from $B$ to 
$E \PAR F$ through $F$ in 
${(\Theta_{S'_{B}})}^{B}$. 
On the other hand, the subpath $\theta_0$ of $\theta$ from $E \PAR F$ 
to $B$ in $\Theta_{S_{B}}$ survives in $\Theta_{S'_{B}}$. 
Therefore $\theta'$ and $\theta_0$ make a cycle in $\Theta_{S'_{B}}$. 
This is a contradiction. 
{\it the end of proof of Subclaim~\ref{subClaimPropEmpireIntersectionEmpty}}\\
Then the following two cases can be considered:
  \begin{enumerate}
  \item The case where there is no path from $A$ to $F$ 
   in $(\Theta_{S'_{B}})^{A}$:\\
   We suppose that there is a unique path $\theta'$ from $A$ to $B$ in $(\Theta_{S'_{B}})^{A}$. 
   Then $\theta'$ does not pass $E$, because if $\theta'$ includes $E$, then 
   $\theta'$ from $A$ to $B$ in $(\Theta_{S'_{B}})^{A}$ survives in $(\Theta_{{S}_{B}})^{A}$ and 
   therefore $\theta$ and $\theta'$ makes a cycle including $B, E \PAR F, E$ in $(\Theta_{{S}_{B}})^{A}$. 
   Moreover $\theta'$ does not pass $E \PAR F$ because if $\theta'$ includes $E \PAR F$, then $\theta'$ also includes $F$, which contradicts the assumption.
   Therefore $\theta'$ survives in $(\Theta_{{S}_{B}})^{A}$. 
   Then $\theta$ and $\theta'$ make a cycle including $A$ and $B$ in 
   $(\Theta_{{S}_{B}})^{A}$. This is a contradiction. 
   Therefore since there is no path $\theta'$ from $A$ to $B$ in $(\Theta_{S'_{B}})^{A}$, we have done. 
   We just let $\Theta^f_{{S}_{B}}$ be $\Theta_{S'_{B}}$.
  \item The case where there is a unique path $\theta'$ from $A$ to $F$ 
   in $(\Theta_{S'_{B}})^{A}$:\\
  Since $F \not\in e_\Theta(A)$, 
  there is a $\PAR$-link $L'$  in $\theta'$ such that exactly one premise and the conclusion of the link are not included 
in $e_\Theta(A)$. 
  Moreover it is obvious that such a $\PAR$-link is unique in $\theta'$. 
  Let $L':\frac{E' \quad F'}{E' \PAR F'}$ be the unique 
  $\PAR$-link. 
Without loss of generality we assume that $\theta'$ passes $E'$, 
$E' \in e_\Theta(A)$, and 
$F' \not\in e_\Theta(A)$. 
Let $S''_{B}$ be the DR-switching $S'_{B}$ except 
$S''_{B}$ chooses the other formula, i.e., $F'$ in $L'$. 
Moreover by the similar discussion to that of $S'_{B}$, 
$S''_{B}$ is a principal switching for $B$.
Then if $S''_{B}$ does not satisfy the condition for ${S}^f_{B}$, then 
we repeat the discussions above to $S''_{B}$. 
Since the number of $\PAR$-links in $\Theta$ is finite, we can eventually find a principal switching 
$S^f_{B}$ for $B$ such that 
there is no path from $A$ to $B$ in 
$(\Theta_{{S}^f_{B}})^{A}$. 
\end{enumerate}
{\it the end of proof of Claim~\ref{claimPropEmpireIntersectionEmpty}}\\
Then since ${S}^f_{B}$ is a principal switching for $B$ in $\Theta$ and $A \not\in e_\Theta(B)$, 
for any formula $C \in e_\Theta(B)$, 
there is no path from $A$ to $C$ in 
$(\Theta_{{S}^f_{B}})^{A}$. 
This means that $C \not\in e_\Theta(A)$. 
This contradicts the assumption $C \in e_\Theta(A) \cap e_\Theta(B)$. 
$\Box$
\end{proof}

The following proposition is given in a stronger form than Lemma 5 of \cite{Gir96} slightly.
\begin{proposition}
\label{propEmpireSubsetNeq}
If $B \not\in e_\Theta(A)$ and 
$A \in e_\Theta(B)$, then 
$A$ is not a conclusion of $e_\Theta(B)$ and 
$e_\Theta(A) \subsetneq e_\Theta(B)$.
\end{proposition}

\begin{proof}
\begin{enumerate}
\item The proof that $A$ is not a conclusion of $e_\Theta(B)$:\\
We suppose that $A$ is a conclusion of $e_\Theta(B)$. 
Let $S$ be a DR-switching. 
Then we claim that $B \in \fml({\Theta_{S}}^{A})$. 
We prove this using case analysis. 
\begin{enumerate}
\item The case where $S$ is a principal switching for $B$:\\ 
By Proposition~\ref{propPrincipalSwitching} 
$\fml({\Theta_{S}}^{B}) = e_\Theta(B)$. 
From assumptions we can easily see that $A$ and $B$ are a leaf or the root in 
the tree ${\Theta_{S}}^{B}$. 
Moreover since $A$ is a conclusion of $e_\Theta(B)$, 
the unique path $\theta$ from $A$ to $B$ in 
${\Theta_{S}}^{B}$ 
immediately above or adjacent to $A$. 
This means that 
$B \in \fml({\Theta_{S}}^{A})$. 
\item The case where $S$ is not a principal switching for $B$:\\ 
Then $A \in e_\Theta(B) \subsetneq \fml({\Theta_S}^{B})$. 
Then there is a unique path $\theta$ from $A$ to $B$ in ${\Theta_S}^{B}$. 
We suppose that $\theta$ passes immediately below $A$.
Then there is the link $L'$ 
whose premise is $A$ and the link $L'$ must be a $\PAR$-link, since 
$A$ is a conclusion of $e_\Theta(B)$. 
Moreover $S$ chooses the premise $A$ in $L'$. 
This means that a formula that is not included in $e_\Theta(B)$ is included in 
$\theta$.
On the other hand let $S_{B}$ be a
principal switching for ${B}$ obtained from $S$ with the minimal effort.
Then for any $\PAR$-link $L_0 \in {\mathbb L}_{e_\Theta(B)}$, 
$S_{B}(L_0) = S(L_0)$ because of the minimal assumption. 
Therefore there is no path $\theta'$ from $A$ to $B$ in 
${\Theta_{S_{B}}}^{B}$ such that
$\theta'$ passes immediately above or adjacent to $A$,
because there is no such path in ${\Theta_{S}}^{B}$. 
Moreover $S_{B}$ chooses another premise other than $A$ 
because $S_{B}$ is a principal switching for $B$. 
Hence there is no path $\theta'$ from $A$ to $B$ 
in $\Theta_{S_{B}}$ such that
$\theta'$ passes immediately below $A$ 
because $S_{B}$ selects the other premise other than $A$ in $L'$.
This means that there is a path $\theta''$ from $B$ to $A$ in 
$\Theta_{S_{B}}$ such that 
$\theta''$ passes immediately below $B$ and immediately above or adjacent to $A$.
This contradicts that $A \in e_\Theta(B)$. 
Therefore $\theta$ passes immediately above or adjacent to $A$. 
This means that $B \in \fml({\Theta_S}^{A})$. 
\end{enumerate}
Therefore, 
\[ B \in \bigcap_{S \, \, \mbox{\scriptsize is a DR-switching for} \, \, \Theta} \fml({\Theta_S}^{A}) 
= e_\Theta(A). 
\]
This contradicts the assumption $B \not\in e_\Theta(A)$. 
\item The proof of $e_\Theta(A) \subsetneq e_\Theta(B)$:\\
Let $S_{A}$ be a principal switching for $A$. 
By Proposition~\ref{propPrincipalSwitching} 
$B \not\in e_\Theta(A) = \fml({(\Theta_{S_{A}})}^{A})$. 
Let $S_{B}$ be a principal switching for $B$ 
obtained from $S_{A}$ by changing $\PAR$-switches with the minimal effort. 
\begin{claim}
\label{claimSubsetNeq1}
Then still $B \not\in \fml({(\Theta_{S_{B}})}^{A})$. 
\end{claim}

{\it Proof of Claim~\ref{claimSubsetNeq1}.} \ \ 
We assume that 
$B \in \fml({(\Theta_{S_{B}})}^{A})$. 
Then there is a unique path $\theta$ from $A$ to $B$ 
in $(\Theta_{S_{B}})^{A}$ 
such that
$\theta$ passes immediately above or adjacent to $A$. 
Since $B \not\in \fml({(\Theta_{S_{A}})}^{A})$ 
and $B \in \fml({(\Theta_{S_{B}})}^{A})$, 
the path $\theta$ must include the conclusion of a $\PAR$-link $L_0$ such that
$S_{A}(L_0) \neq S_{B}(L_0)$. 
On the other hand, by the minimal assumption about the change from $S_{A}$ to $S_{B}$,
the conclusion of $L_0$ is not included in $e_\Theta(B)$.
Moreover since $A \in e_\Theta(B) = {(\Theta_{S_{B}})}^{B}$, 
there is a path $\theta'$ from $A$ to $B$ 
in ${(\Theta_{S_{B}})}^{B}$
such that 
all the $\PAR$-formulas in $\theta'$ are included in ${e_\Theta(B)}$.
Therefore since 
these two paths $\theta$ and $\theta'$ from $A$ to $B$ in $\Theta_{S_{B}}$ are different, $\theta$ and $\theta'$ 
make a cycle in $\Theta_{S_{B}}$. 
This is a contradiction. 
\ {\it the end of proof of Claim~\ref{claimSubsetNeq1}}\\
Then we can prove the following.

\begin{claim}
\label{claimSubsetNeq2}
$\fml({(\Theta_{S_{B}})}^{A}) \subseteq 
\fml({(\Theta_{S_{B}})}^{B})$
\end{claim}

{\it Proof of Claim~\ref{claimSubsetNeq2}.} \ \ 
We assume that 
there is a formula $C \in \fml({(\Theta_{S_{B}})}^{A})$,
but $C \not\in \fml({(\Theta_{S_{B}})}^{B})$. 
Since $A \in e_\Theta(B) = \fml({(\Theta_{S_{B}})}^{B})$ and $S_{B}$ is a principal switching for $B$, 
the unique path $\pi'$ from $A$ to $C$ in 
$\Theta_{S_{B}}$ must include $B$ in order to 
go out from $e_\Theta(B) = \fml({(\Theta_{S_{B}})}^{B})$.
On the other hand, since $C \in \fml({(\Theta_{S_{B}})}^{A})$, 
there is the unique path $\pi''$ from $A$ to $C$ in 
${(\Theta_{S_{B}})}^{A}$ such that 
$\pi''$ passes immediately above or adjacent to $A$. 
By uniqueness $\pi'$ and $\pi''$ coincide in $\Theta_{S_{B}}$. 
Therefore there is a subpath $\pi'_0$ of $\pi'$ from $A$ to $B$ such that $\pi'_0$ passes immediately above or adjacent to both 
$A$ and $B$. 
Hence we can derive 
$B \in \fml({(\Theta_{S_{B}})}^{A})$. 
This contradicts $B \not\in \fml({(\Theta_{S_{B}})}^{A})$.
\ {\it the end of proof of Claim~\ref{claimSubsetNeq2}}\\
Therefore 
$e_\Theta(A) \subseteq \fml({(\Theta_{S_{B}})}^{A}) \subseteq 
\fml({(\Theta_{S_{B}})}^{B}) = e_\Theta(B)$.
$\Box$
\end{enumerate}
\end{proof}

\begin{proposition}
\label{propEmpireTensPremise}
Let $\Theta$ be an MLL proof net including $\TENS$-link 
$L:\frac{A \quad B}{A \TENS B}$. 
Then $e_\Theta(A) \cap e_\Theta(B) = \emptyset$.
\end{proposition}

\begin{proof}
We assume $e_\Theta(A) \cap e_\Theta(B) \neq \emptyset$.
Then $A \TENS B \not\in e_\Theta(A) \cap e_\Theta(B)$.
Otherwise, there is a DR-switching $S$ for $\Theta$  
such that $\Theta_S$ has a cycle including $A$ and $A \TENS B$. 
Therefore there is a formula $C$ such that $C \in e_\Theta(A) \cap e_\Theta(B)$ and $k \neq \ell$. 
Then when we consider $e_\Theta(A \TENS B)$, 
we can easily see that there is an arbitrary DR-switching $S$ for $\Theta$ such that $\Theta_S$ has a cycle including $C$ and $A \PAR B$, 
since there is a unique path from $A$ to $C$ 
in $\Theta_S^{A}$ and
there is also the unique path from $B$ to $C$
in ${\Theta_S}^{B}$. 
This is a contradiction. 
$\Box$
\end{proof}

\begin{proposition}
\label{propEmpireParPremise}
Let $\Theta$ be an MLL proof net including $\PAR$-link 
$L:\frac{A \quad B}{A \PAR B}$. 
Then $e_\Theta(A) = e_\Theta(B)$.
\end{proposition}

\begin{proof}
\begin{claim}
\label{claimPropEmpireParPremise}
$e_\Theta(A) \cap e_\Theta(B) \neq \emptyset$
\end{claim}
{\it Proof of Claim~\ref{claimPropEmpireParPremise}.} \ \ 
We assume that
$e_\Theta(A) \cap e_\Theta(B) = \emptyset$. 
We take a principal switching $S_{B}$ for $B$. 
Then 
there is no path from $A$ to $B$ 
in $\Theta_{S_{B}}$. 
In order to prove this, we assume that 
there is a path $\theta$ from $A$ to $B$ 
in $\Theta_{S_{B}}$. 
The path $\theta$ does not pass immediately below $B$.
If so,  since $S_{B}$ is a principal switching for $B$, $S_{B}$ selects $B$ in the $\PAR$-link $L$. 
Therefore, $\theta$ passes immediately above or adjacent to $A$. 
Moreover by the assumption, $\theta$ includes the subpath $A \PAR B, B$. 
Then let $S_A$ be $S_B$ except that $S_A$ chooses $A$ in $L$.
Then $S_A(\Theta)$ has a cycle including the subpath of $\theta$ from $A$ to $A \PAR B$ and the path $A \PAR B, A$. 
Therefore the path $\theta$ does not pass immediately below $B$.
On the other hand, the path $\theta$ does not pass immediately above or adjacent to $B$
because $A \not\in e_\Theta(B)$ (since $e_\Theta(A) \cap e_\Theta(B) = \emptyset$)
and $\fml((\Theta_{S_{B}})^{B}) = e_\Theta(B)$. 
Therefore $\Theta_{S_{B}}$ is disconnected. 
This is a contradiction. 
\ {\it the end of proof of Claim~\ref{claimPropEmpireParPremise}}\\
Then by Proposition~\ref{propEmpireIntersectionEmpty}, 
$B \in e_\Theta(A)$ or 
$A \in e_\Theta(B)$.
\begin{enumerate}
\item The case where $B \in e_\Theta(A)$ 
and $A \in e_\Theta(B)$:\\
It is obvious that $A \PAR B \not\in e_\Theta(A)$, 
since otherwise we can easily find a DR-switching $S$ such that $\Theta_S$ has a cycle including 
$A$ and 
$A \PAR B$. Similarly $A \PAR B \not\in e_\Theta(B)$.
So $B$ is a conclusion of $e_\Theta(A)$ and 
$A$ is a conclusion of $e_\Theta(B)$. \\
Let $S_{B}$ be a principal switching for $B$.
In addition, let $S_{A}$ be a principal switching for $A$ 
obtained from $S_{B}$ by changing $\PAR$-switches with the minimal effort. 
Then the following claim holds.
\begin{claim}
\label{claimCaseB-eA-and-A-eB-1}
Let $C \in \fml({(\Theta_{S_{B}})}^{B})$ 
and 
$\theta$ be a unique path from $A$ to $C$ 
in ${(\Theta_{S_{B}})}^{B}$. 
Then each formula in $\theta$ is included in 
${(\Theta_{S_{A}})}^{A}$.
\end{claim}

{\it Proof of Claim~\ref{claimCaseB-eA-and-A-eB-1}.} \ \ 
At first we note that 
$\theta$ passes immediately above or adjacent to $A$ because
$S_{B}$ selects $B$ in the $\PAR$-link $L$.
We assume that the statement does not hold.
Then without loss of generality, there is a subpath $E, E \PAR F$ 
in $\theta$ such that 
the subpath of $\theta$ from $A$ to $E$ in ${(\Theta_{S_{B}})}^{B}$ survives in 
${(\Theta_{S_{A}})}^{A}$
and
$E \in \fml({(\Theta_{S_{A}})}^{A})$,
but $E \PAR F \not\in \fml({(\Theta_{S_{A}})}^{A})$.
Moreover, 
since $S_{A}$ is principal for $A$, 
there is a path $\pi$ in $\Theta_{S_{A}}$ from $A$ 
to $F$ such that 
$\pi$ passes immediately below $A$ in $\Theta_{S_{A}}$.
Then each formula in $\pi$ except $A$ 
does not belong to $\fml({(\Theta_{S_{B}})}^{B})$.
In fact, let $G$ be the first formula in $\pi$ except $A$ 
such that $G \in \fml({(\Theta_{S_{B}})}^{B})$. 
Then the subpath $\pi'$ of $\pi$ from $A \PAR B$ to $G$ 
in $\Theta_{S_{A}}$
survives in 
$\Theta_{S_{B}}$. 
On the other hand, 
since $G \in \fml({(\Theta_{S_{B}})}^{B})$, 
there is a unique path $\xi$ from $B$ to $G$ in 
${(\Theta_{S_{B}})}^{B}$ 
such that $\xi$ passes immediately above or adjacent to $B$. 
Then since $S_{B}$ selects $B$ in the $\PAR$-link $L$, 
$\pi'$ and $\xi$ makes a cycle in $\Theta_{S_{B}}$. 
This is a contradiction. 
Therefore, each formula in $\pi$ except $A$ does not belong to $\fml({(\Theta_{S_{B}})}^{B})$.
But $F \in \fml({(\Theta_{S_{B}})}^{B})$ 
because $E \PAR F$ belongs to $\theta$ and 
$\theta$ is included in ${(\Theta_{S_{B}})}^{B}$. 
This is a contradiction. 
\ {\it the end of proof of Claim~\ref{claimCaseB-eA-and-A-eB-1}}\\
Since $S_{B}$ (resp. $S_{A}$) 
is a principal switching for $B$ (resp. $A$), 
Claim~\ref{claimCaseB-eA-and-A-eB-1} means 
$e_\Theta(B) \subseteq e_\Theta(A)$.
Similarly we can prove $e_\Theta(A) \subseteq e_\Theta(B)$.
So $e_\Theta(A) = e_\Theta(B)$.
\item The case where $B \not\in e_\Theta(A)$ 
and $A \in e_\Theta(B)$:\\
Then by Proposition~\ref{propEmpireSubsetNeq}, 
$e_\Theta(A) \subsetneq e_\Theta(B)$ and $A$ is not a conclusion of $e_\Theta(B)$. 
But this implies $A \PAR B \in e_\Theta(B)$, which contradicts 
the definition of empires. Therefore this case never happens. 
\item The case where $B \in e_\Theta(A)$ 
and $A \not\in e_\Theta(B)$:\\
Similar to the case immediately above. 
$\Box$
\end{enumerate}
\end{proof}

The next goal is to prove Splitting lemma (Lemma~\ref{lemmaSplitting}). 
In order to do that, we introduce a strict partial order on $\TENS$-formulas in a MLL proof net.

\begin{definition}
\label{defStrictPartialOrderOfTENS}
Let $\Theta$ be an MLL proof net. 
Let $L:\frac{A \quad B}{A \TENS B}$ 
and $L':\frac{A' \quad B'}{A' \TENS B'}$ 
be $\TENS$-links in $\Theta$. 
Then, 
\[ A \TENS B < A' \TENS B' \, \, \mbox{iff} \, \, 
e_{\Theta_0}(A \TENS B) \subseteq e_{\Theta_0}(A') \vee 
e_{\Theta_0}(A \TENS B) \subseteq e_{\Theta_0}(B')
\]
\end{definition}

\begin{proposition}
\label{propStrictPartialOrderOfTENS}
$<$ is a strict partial order. 
\end{proposition}

\begin{proof}
\begin{itemize}
\item transitivity:\\
We assume that $A \TENS B < A' \TENS B'$  and 
$A' \TENS B' < A'' \TENS B''$. By definition,
$(e_{\Theta_0}(A \TENS B) \subseteq e_{\Theta_0}(A') \vee 
e_{\Theta_0}(A \TENS B) \subseteq e_{\Theta_0}(B')) \wedge
(e_{\Theta_0}(A' \TENS B') \subseteq e_{\Theta_0}(A'') \vee 
e_{\Theta_0}(A' \TENS B') \subseteq e_{\Theta_0}(B''))$.
We only consider the case where 
$e_{\Theta_0}(A \TENS B) \subseteq e_{\Theta_0}(A') \wedge e_{\Theta_0}(A' \TENS B') \subseteq e_{\Theta_0}(B'')$ because the other three cases are similar. 
Since $e_{\Theta_0}(A \TENS B) \subseteq e_{\Theta_0}(A')$ and 
$e_{\Theta_0}(A') \subseteq e_{\Theta_0}(A' \TENS B')$, 
we obtain $e_{\Theta_0}(A \TENS B) \subseteq e_{\Theta_0}(A' \TENS B')$. 
Therefore from $e_{\Theta_0}(A' \TENS B') \subseteq e_{\Theta_0}(B'')$, 
we obtain $e_{\Theta_0}(A \TENS B) \subseteq e_{\Theta_0}(B'')$. 
So, $A \TENS B < A'' \TENS B''$. 
\item irreflexivity:\\
We assume that $A \TENS B < A \TENS B$. 
Then by definition $e_{\Theta_0}(A \TENS B) \subseteq e_{\Theta_0}(A) \vee 
e_{\Theta_0}(A \TENS B) \subseteq e_{\Theta_0}(B)$. 
We only consider the case where $e_{\Theta_0}(A \TENS B) \subseteq e_{\Theta_0}(A)$, because 
the other case is similar. 
Then $B \in e_{\Theta_0}(A \TENS B) \subseteq e_{\Theta_0}(A)$ 
and $B \in e_{\Theta_0}(B)$. 
So $e_{\Theta_0}(A) \cap e_{\Theta_0}(B) \neq \emptyset$. 
From Proposition~\ref{propEmpireTensPremise} We derive a contradiction. 
$\Box$
\end{itemize}
\end{proof}

\begin{lemma}[Splitting Lemma]
\label{lemmaSplitting}
Let $\Theta$ be an MLL proof net whose conclusions does not include any $\PAR$-formulas. 
Then there is a conclusion $L:\frac{A \quad B}{A \TENS B}$ in $\Theta$
such that $\fml(\Theta) = \{ A \TENS B \} \uplus e_\Theta(A) \uplus e_\Theta(B)$. 
\end{lemma}

\begin{proof}
Let $T = \{ A_1 \TENS B_1, \ldots, A_\ell \TENS B_\ell \}$ be the conclusions in $\Theta$
that are a $\TENS$-formula. 
Then let $\ell_0 \, (1 \le \ell_0 \le \ell)$ be an index such that 
$A_{\ell_0} \TENS B_{\ell_0}$ is a maximal element in $T$ w.r.t the strict partial order $<$. 
We can always find the index by the finiteness of $\Theta$. 
We claim that $A_{\ell_0} \TENS B_{\ell_0}$ is $A \TENS B$ of the the statement. 
We assume that $\ell_0$ is not. 
Then without loss of generality, there is a conclusion $C$ of 
$e_\Theta(A_{\ell_0})$ such that $C$ is not a conclusion of $\Theta$.
Then without loss of generality there is an index $\ell' \, (1 \le \ell' \le \ell)$ such that 
$C$ is hereditarily above $B_{\ell'}$. 
Hence by Proposition~\ref{propEmpireIncudinglPremises}, 
$C \in e_\Theta(B_{\ell'})$. 
Moreover, from the definition of empires, 
$B_{\ell'} \not\in e_\Theta(A_{\ell_0})$. 
Then, by Proposition~\ref{propEmpireIntersectionEmpty}, 
$A_{\ell_0} \in e_\Theta(B_{\ell'})$. 
Hence by Proposition~\ref{propEmpireSubsetNeq}, 
$e_\Theta(A_{\ell_0}) \subsetneq e_\Theta(B_{\ell'})$. 
So, since 
$e_\Theta(A_{\ell_0} \TENS B_{\ell_0}) 
\subsetneq e_\Theta(B_{\ell'})$. 
Hence $A_{\ell_0} \TENS B_{\ell_0} < A_{\ell'} \TENS B_{\ell'}$. 
This contradicts the maximality of $A_{\ell_0} \TENS B_{\ell_0}$ w.r.t $<$ over $T$. 
$\Box$
\end{proof}

\section{Proof of Proposition~\ref{propPNGraphAutoUnique}}
\label{secPNGraphAutoUnique}
\begin{flushleft}
{\it Proof of Proposition~\ref{propPNGraphAutoUnique}} \ 
We prove this proposition by induction on the number of the links in $\Theta$. 
Before that, we prove the following claim.

\begin{claim}
\label{claimDiffAutomorAllDiff}
Let $\langle h_V, h_E \rangle$ be an other graph automorphism on 
$G^{\stripPT}(\Theta)$ than $\langle \id_V, \id_E \rangle$. 
Then $\forall v \in V. h_V(v) \neq v$. 
\end{claim}
{\it proof of Claim~\ref{claimDiffAutomorAllDiff}: \ }
We assume $h_V = \id_V$. 
Since $\langle h_V, h_E \rangle \neq \langle \id_V, \id_E \rangle$,
there is $e_0 \in E$ such that $e_0 = \id_E(e_0) \neq h_E(e_0)$. 
On the other hand, since $\langle h_V, h_E \rangle$ is a graph automorphism, 
$\ell_E(e_0) = \ell_E(h_E(e_0)) \in \{ {\bf L}, {\bf R}, {\bf ID} \}$. 
Therefore the link $L$ that induces $e_0$ is different from the link $L'$ that induces $h_E(e_0)$. 
Then since (a) two different links does not share the same formula 
except that the formula is one premise of the one link and one conclusion of the other link,
but (b) $\src(e_0)$ is a conclusion (resp. premise) of $L$ iff $\src(h_E(e_0))$ 
is a conclusion (resp. premise) of $L'$, 
hence, $h_V(\src(e_0)) = \src(h_E(e_0)) \neq \src(e_0)$. 
Therefore $h_V \neq \id_V$. \\
So, there is $v_0 \in V$ such that $v_0 = \id_V(v_0) \neq h_V(v_0)$. 
The the following subclaim holds.
\begin{subclaim}
\label{subclaimDiffAutomorAllDiff}
For any $e \in E$ and $v \in V$, 
if $\src(e) = v_0$ and $\tgt(e) = v$, or $\src(e) = v$ and $\tgt(e) = v_0$, then 
$v \neq h_V(v)$. 
\end{subclaim}
{\it proof of Subclaim~\ref{subclaimDiffAutomorAllDiff}: \ }
We only consider the case where $\src(e) = v_0$ and $\tgt(e) = v$, because the other case is similar. 
Since $v_0 \neq h(v_0)$ and $\ell_V^{\stripPT}(v_0) = \ell_V^{\stripPT}(h_V(v_0))$, hence, 
$e \neq h_E(e)$. 
Then since $\ell_E(e) = \ell_E(h_E(e)) \in \{ {\bf L}, {\bf R}, {\bf ID} \}$, 
by the same discussion above, 
we can derive $v = \tgt(e) \neq h_V(\tgt(e)) = h_V(v)$. 
\ {\it \ the end of the proof of Subclaim~\ref{subclaimDiffAutomorAllDiff}} \\
Since $\Theta$ is an MLL proof net, 
starting from $v_0 \in V$, we can reach any $v \in V$ by 
moving from a node $v_1 \in V$ to another node $v_2 \in V$ repeatedly such that 
$v_1$ and $v_2$ are a premise or a conclusion of the same link. 
Then through the travelling, by applying the subclaim, we can derive the claim. 
\ {\it \ the end of the proof of Claim~\ref{claimDiffAutomorAllDiff}} 
Then we prove the proposition using the claim above. 
\begin{enumerate}
\item [1.] The case where $\Theta$ consists of exactly one ID-link $\overline{p \quad p^\bot}$:\\
  It is obvious that the identity map is the only graph automorphism on $G^{\stripPT}(\Theta)$. 
\item [2.] The case where there is a $\PAR$-formula $\langle A \PAR B, k_1 \rangle$ among the conclusions in $\Theta$:\\
  Let $\langle h_V, h_E \rangle$ be an other graph automorphism 
  on $G^{\stripPT}(\Theta)$ than $\langle \id_V, \id_E \rangle$. 
  By Claim~\ref{claimDiffAutomorAllDiff}, 
  $\Theta$ must have a conclusion $\langle A \PAR B, k_2 \rangle$ such that 
  $k_1 \neq k_2$, $h_V(k_1) = k_2$, and $h_V(k_2) = k_1$. 
  Let $\Theta_0$ be the proof net obtained from $\Theta$ deleting the two $\PAR$-links associated with 
  $\langle A \PAR B, k_1 \rangle$ and $\langle A \PAR B, k_2 \rangle$ 
  (let the two $\PAR$-links be 
  $L_{\PAR 1}: \frac{\langle A, i_1 \rangle \quad \langle B, j_1 \rangle}{\langle A \PAR B, k_1 \rangle}$ and 
  $L_{\PAR 2}: \frac{\langle A, i_2 \rangle \quad \langle B, j_2 \rangle}{\langle A \PAR B, k_2 \rangle}$ respectively). 
  We apply inductive hypothesis to $\Theta_0$. 
  Then the only graph automorphism on $G^{\stripPT}(\Theta_0)(= \langle V_0, E_0, \ell_{V_0}^{\stripPT}, \ell_{E_0} \rangle)$  is $\langle \id_{V_0}, \id_{E_0} \rangle$. 
  Therefore $\langle h_V, h_E \rangle$ must be an extension of $\langle \id_{V_0}, \id_{E_0} \rangle$. 
  But it is impossible, because since $h_V(k_1) = k_2$, and $h_V(k_2) = k_1$, 
  we must have $h_V(i_1) = i_2$, $h_V(i_2) = i_1$, $h_V(j_1) = j_2$, and $h_V(j_2) = j_1$. 
\item [3.] The case where there is no $\PAR$-formula among the conclusions in $\Theta$:\\
  In this case, by applying Lemma~\ref{lemmaSplitting} (Appendix~\ref{secEmpBasicProp}) to $\Theta$ we can find 
  $\langle A_1 \TENS B_1, k_1 \rangle$ such that 
  $\fml(\Theta) = \{ \langle A_1 \TENS B_1, k_1 \rangle \} \uplus e_{\Theta}(\langle A, i_1 \rangle) \uplus  e_{\Theta}(\langle B, j_1 \rangle)$.
  Let $\langle h_V, h_E \rangle$ be an other graph automorphism 
  on $G^{\stripPT}(\Theta)$ than $\langle \id_V, \id_E \rangle$. 
  By Claim~\ref{claimDiffAutomorAllDiff}, 
  $\Theta$ must have a conclusion $\langle A \TENS B, k_2 \rangle$ such that 
  $k_1 \neq k_2$, $h_V(k_1) = k_2$, and $h_V(k_2) = k_1$. 
  Moreover by symmetry, 
  we must have $\fml(\Theta) = \{ \langle A \TENS B, k_2 \rangle \} \uplus e_{\Theta}(\langle A, i_2 \rangle) \uplus  e_{\Theta}(\langle B, j_2 \rangle)$. 
  Moreover by symmetry, it is enough to consider the following two cases. 
  \begin{enumerate}
  \item [(a)] The case where $\fml(\Theta) = \{ \langle A \TENS B, k_1 \rangle, \langle A \TENS B, k_2 \rangle \} \uplus e_{\Theta}(\langle A, i_1 \rangle) \uplus e_{\Theta}(\langle A, i_2 \rangle) 
                       \uplus \bigl( e_{\Theta}(\langle B, j_1 \rangle) \cap e_{\Theta}(\langle B, j_2 \rangle) \bigr)$
  \item [(b)] The case where $\fml(\Theta) = \{ \langle A \TENS B, k_1 \rangle, \langle A \TENS B, k_2 \rangle \} \uplus e_{\Theta}(\langle B, j_1 \rangle) \uplus e_{\Theta}(\langle B, j_2 \rangle) 
                       \uplus \bigl( e_{\Theta}(\langle A, i_1 \rangle) \cap e_{\Theta}(\langle A, i_2 \rangle) \bigr)$
  \end{enumerate}
  We only consider the case (a) because the case (b) is similar. 
  Then let $\Theta_0$ be the proof net whose formulas are 
  $e_{\Theta}(\langle B, j_1 \rangle) \cap e_{\Theta}(\langle B, j_2 \rangle)$. 
  We apply inductive hypothesis to $\Theta_0$. 
  Then the only graph automorphism on $G^{\stripPT}(\Theta_0)(= \langle V_0, E_0, \ell_{V_0}^{\stripPT}, \ell_{E_0} \rangle)$  is $\langle \id_{V_0}, \id_{E_0} \rangle$. 
  Therefore $\langle h_V, h_E \rangle$ must be an extension of $\langle \id_{V_0}, \id_{E_0} \rangle$. 
  But it is impossible, because since $h_V(k_1) = k_2$, and $h_V(k_2) = k_1$, 
  we must have $h_V(i_1) = i_2$, $h_V(i_2) = i_1$, $h_V(j_1) = j_2$, and $h_V(j_2) = j_1$. 
$\Box$
\end{enumerate}
\end{flushleft}

\section{Proof of Theorem~\ref{mainTheorem1}}
\label{secProofOfMainTheorem1}
\begin{flushleft}{\it Proof of Theorem~\ref{mainTheorem1}.} \ \ 
At first we fix our notation. 
Let $\Theta'$ be $\tpex(\Theta, L_{1 \TENS}, L_{\PAR2})$ 
and $L'_{1\PAR}$ and $L'_{2\TENS}$ be 
$\frac{A \quad B}{A \PAR B}$ and
$\frac{C \quad D}{C \TENS D}$ respectively. \\
$\bullet$ {\bf If part}
\begin{enumerate}
\item [1.] The case where $C$ is a conclusion of $e_\Theta(A)$ and 
$D$ is a conclusion of $e_\Theta(B)$:\\
Let $S'$ be a DR-switching for $\Theta'$. We assume that $\Theta'_{S'}$ has a cycle or is disconnected.
\begin{enumerate}
\item [(a)] The case where $S'$ selects $A$ in ${L'}_{1\PAR}$:\\
By the assumption on $\Theta'_{S'}$, 
(i) there is a cycle including $C \TENS D$ in $\Theta'_{S'}$ or
(ii) $A$ and $B$ are disconnected in $\Theta'_{S'}$.
Then let $S$ be the DR-switching for $\Theta$ such that 
$S$ is $S'$ except that 
$S$ chooses the left or the right premise of $L_{2\PAR}$ 
and the domain of $S$ does not include $L'_{1\PAR}$.
Then there are two unique paths $\theta_1$ and $\theta_2$ in $\Theta_S$ 
from $A$ to $C$ and 
from $B$ to $D$ respectively. 
From our assumption about $C$ and $D$, we can easily see that all the indexed formulas in $\theta_1$ and $\theta_2$ are 
included in $e_\Theta(A)$ and $e_\Theta(B)$ respectively.
In particular, 
\begin{enumerate}
\item [$\circ$] $\theta_1$ passes immediately above or adjacent to both 
$A$ and $C$, and 
\item [$\circ$] $\theta_2$ passes immediately above or adjacent to both
$B$ and $D$. 
\end{enumerate} 
Moreover, by our assumption and Proposition~\ref{propEmpireTensPremise} 
we obtain $e_\Theta(A) \cap e_\Theta(B) = \emptyset$. 
Therefore if we consider $\theta_1$ and $\theta_2$ as two sets of indexed formulas, 
$\theta_1$ and $\theta_2$ are disjoint. Moreover, 
two paths $\theta_1$ and $\theta_2$ in $\Theta_S$ are preserved in $\Theta'_{S'}$
because $\theta_1$ (resp. $\theta_2$) includes neither $A \TENS B$ nor $C \PAR D$. 
Hence if we let $(\theta_2)^{r}$ be the reverse of $\theta_2$, then 
$\theta_1, C \TENS D, (\theta_2)^{r}$ is the unique path 
from $A$ to $B$ in $\Theta'_{S'}$. Hence the case (ii) is impossible. 
So the case (i) holds. \\
If $\Theta'_{S'}$ has a cycle $\pi$, then one of the following conditions must be satisfied:
\begin{enumerate}
\item [(a-1)] The case where the cycle $\pi$ in $\Theta'_{S'}$ includes 
  $C, C \TENS D, D$:\\
  Since $C \in e_\Theta(A)$,
        $D \in e_\Theta(B)$
        and $e_\Theta(A) \cap e_\Theta(B) = \emptyset$, $\pi$ must include at least one indexed formula from each of the following three types of indexed formulas except $C, C \TENS D, D$:
  (I) indexed formulas from $e_\Theta(A)$ 
           different from $A$,  
  (II) indexed formulas from $e_\Theta(B)$
           different from $B$, and 
  (III) indexed formulas that are not included in 
           $e_\Theta(A) \cup e_\Theta(B)$. 
  Let $E$ be an indexed formula of the type (I) 
  that is included in $\pi$ and 
      $F$ be an indexed formula of the type (II) that is included in $\pi$. 
  Then there is a path $\tau_1$ from $A$ to $E$ in $\Theta'_{S'}$ 
  such that all the indexed formulas in $\tau_1$ are included in 
  $e_\Theta(A)$ and $\tau_1$ passes immediately above or adjacent to $A$. 
Similarly, 
  there is a path $\tau_2$ from $B$ to $F$ in $\Theta'_{S'}$ 
  such that all the indexed formulas in $\tau_2$ are included in 
  $e_\Theta(B)$ and $\tau_2$ passes immediately above or adjacent to $B$.  On the other hand since $\pi$ has indexed formulas of type (III), 
  there is the subpath $\pi'$ of $\pi$ from $E$ to $F$ 
  such that $\pi'$ includes at least one indexed formula that is not included in 
  $e_\Theta(A) \cup e_\Theta(B)$. \\
  Since $\Theta$ is an MLL proof net, $\Theta_S$ must be acyclic and connected. 
  But there is the cycle $A \TENS B, \tau_1, \pi', (\tau_2)^r, A \TENS B$ 
  in $\Theta_S$. 
  This is a contradiction.
\item [(a-2)] The case where the cycle $\pi$ in $\Theta'_{S'}$ includes 
  $C$ and $C \TENS D$, but does not include $D$:\\
  In this case 
  there is the subpath $\pi_0$ of $\pi$ from $C$ to $C \TENS D$ in $\Theta'_{S'}$ 
  such that $\pi_0$ passes immediately above or adjacent to $C$ and 
  immediately below $C \TENS D$. 
  We let the DR-switching $S$ for $\Theta$ select $C$ in $L_{2 \PAR}$. 
  Since $\Theta$ is an MLL proof net, $\Theta_S$ must be acyclic and connected. 
  But since $\pi_0$ in $\Theta'_{S'}$ survives in $\Theta_S$, $\Theta_S$ has a cycle. 
  This is a contradiction. 
\item [(a-3)] The case where the cycle $\pi$ in $\Theta'_{S'}$ includes 
  $D$ and $C \TENS D$, but does not include $C$:\\
  Similar to the case immediately above except that 
  we let the DR-switching $S$ for $\Theta$ select $D$ in $L_{2 \PAR}$. 
\end{enumerate}
\item [(b)] The case where $S'$ selects $B$ in ${L'}_{1 \PAR}$:\\
  Similar to the case above.
\end{enumerate}
\item [2.] The case where $D$ is a conclusion of $e_\Theta(A)$ and 
$C$ is a conclusion of $e_\Theta(B)$:\\
Similar to the case above.
\end{enumerate}
$\bullet$ {\bf Only-if part}\\
We suppose that $\Theta$ and $\Theta' (=\tpex(\Theta, L_{1 \TENS}, L_{\PAR2}))$ are proof nets, 
but neither (1) nor (2) of the statement of the theorem holds.
Then we derive a contradiction. 
Basically we find a DR-switching $S'$ for $\Theta'$ such that $S'(\Theta')$ has a cycle. 
We prove this by case analysis.
\begin{enumerate}
\item [1.] The case where $C \PAR D \in e_\Theta(A)$:\\
By Proposition~\ref{propEmpireIncudinglPremises}, $C \in e_\Theta(A)$ and 
$D \in e_\Theta(A)$. 
Let $S$ be a principal DR-switching for $A$ in $\Theta$. Without loss of generality we assume that 
$S$ selects $C$ in $L_{2 \PAR}$. 
Since $\Theta_S$ is acyclic and connected, 
there are two unique paths $\theta_1$ from $A$ to $C$ 
and $\theta_2$ from $A$ to $D$ in ${(\Theta_S)}^A$ 
such that 
both $\theta_1$ and $\theta_2$ pass immediately above or adjacent to $A$. 
Moreover, since $e_\Theta(A) \cap e_\Theta(B) = \emptyset$ 
and all the formulas in  $\theta_1$ and $\theta_2$ are included in $e_\Theta(A)$, 
neither $\theta_1$ nor $\theta_2$ includes $B$. 
We have two cases.
\begin{enumerate}
\item The case where both $\theta_1$ and $\theta_2$ pass $C \PAR D$:\\
In this case, both $\theta_1$ and $\theta_2$ pass immediately below $C \PAR D$.
Otherwise, let $S_0$ be $S$ except $S_0$ selects $D$ in $L_{2 \PAR}$. 
Then $S_0(\Theta)$ has a cycle including the subpath of $\theta_1$ from $A$ to $C \PAR D$, 
the path $C \PAR D, D$, and ${(\theta_2)}^r$ from $D$ to $A$. This is a contradiction.
Therefore since $S$ selects $C$ in $L_{2 \PAR}$, 
$\theta_1$ from $A$ to $C$ is a subpath of $\theta_2$ from $A$ to $D$ in ${S(\Theta)}^A$.
Hence $\theta_2$ has the subpath $\theta_{21}$ from $C \PAR D$ to $D$ such that 
$\theta_{21}$ passes immediately above or adjacent to $C \PAR D$. 
Let $S'$ be $S$ except that the $\PAR$-switch for $L_{2 \PAR}$ is deleted and
the $\PAR$-switch for $L'_{1 \PAR}$ selects $A$ or $B$. 
Then $S'$ is a DR-switching for $\Theta'$ and $S'(\Theta')$ includes a cycle $\theta_{21}, C \TENS D$. 
\item Otherwise:\\
In this case, neither $\theta_1$ nor $\theta_2$ includes $C \PAR D$ (otherwise, 
we have a cycle including $C, C \PAR D$ in $\Theta_{S}$ or
when we let $S_{\bf R}$ be the DR-switching obtained from $S$ by selecting $D$ in $L_{2 \PAR}$, 
we have a cycle including $D, C \PAR D$ in $\Theta_{S_{\bf R}}$). 
Therefore $\theta_1$ (resp. $\theta_2$) passes immediately above or adjacent to $C$ (resp. $D$).
Then let $S'$ be $S$ except that the $\PAR$-switch for $L_{2 \PAR}$ is deleted and
the $\PAR$-switch for $L'_{1 \PAR}$ selects $A$ or $B$. 
Since $e_\Theta(A) \cap e_\Theta(B) = \emptyset$ 
and all the formulas in  $\theta_1$ and $\theta_2$ are included in $e_\Theta(A)$, 
both $\theta_1$ and $\theta_2$ in $S(\Theta)$ survive in $S'(\Theta')$.
Then we find a cycle $\theta_1, C \TENS D, {(\theta_2)}^r$ in $S'(\Theta')$. 
\end{enumerate}
\item [2.] The case where $C \PAR D \in e_\Theta(B)$:\\
Similar to the case above.
\item [3.] The case where $C \PAR D \not\in e_\Theta(A)$ and $C \PAR D \not\in e_\Theta(B)$:\\
Moreover we divide the case into two cases.
\begin{enumerate}
\item [(a)] The case where $e_\Theta(A \TENS B) \cap e_\Theta(C \PAR D) = \emptyset$:\\
Let $S_{\bf L}$ be a DR-switching for $\Theta$ selecting $C$ in $L_{2 \PAR}$. 
Then there is the unique path $\theta$ from $D$ to $C \PAR D$ in 
$\Theta_{S_{\bf L}}$ with length $>1$. 
Let $S'_{\bf L}$ be $S_{\bf L}$ except that 
the $\PAR$-switch for $L_{2 \PAR}$ is deleted and
the $\PAR$-switch for ${L'}_{1 \PAR}$ selects $A$ (or $B$). 
Then $\Theta'_{S'_{\bf L}}$ has a cycle 
$\theta, D$. 
This is a contradiction.
\item [(b)] The case where $e_\Theta(A \TENS B) \cap e_\Theta(C \PAR D) \neq \emptyset$:\\
Then by Proposition~\ref{propEmpireIntersectionEmpty},
$C \PAR D \in e_\Theta(A \TENS B)$ 
or $A \TENS B \in  e_\Theta(C \PAR D)$. 

\begin{enumerate}
\item [(b-1)] The case where $C \PAR D \in e_\Theta(A \TENS B)$: \\
Since neither (1) nor (2) of the statement of the theorem holds, 
one of the following four cases must hold. 
\begin{enumerate}
\item [(b-1-1)] The case where neither $C$ nor $D$ is a 
            conclusion of $e_\Theta(A)$:\\
            In this case, since $C \PAR D \not\in e_\Theta(A)$, 
            $C \not\in e_\Theta(A)$ and 
            $D \not\in e_\Theta(A)$.
	    Let $S_{{\bf L} A}$ be a principal switching for $A$ 
            and $\Theta$ such that $S_{{\bf L} A}$ selects $C$ in
            $L_{2 \PAR}$. 
            Then there are two unique paths $\theta_1$ from $A$ to 
            $C$ and $\theta_2$ from $A$ to $D$
            in $\Theta_{S_{{\bf L} A}}$ 
            such that both $\theta_1$ and $\theta_2$ pass immediately below $A$. 
           Let $S'$ be $S_{{\bf L} A}$ except that 
the $\PAR$-switch for $L_{2 \PAR}$ is deleted and
the $\PAR$-switch for ${L'}_{1 \PAR}$ selects $A$ (or $B$). 
Then $\Theta'_{S'}$ has a cycle $\theta_1, C \TENS D, {(\theta_2)}^r$. 
This is a contradiction. 
\item [(b-1-2)] The case where $C$ is neither a conclusion of $e_\Theta(A)$ 
            nor a conclusion of $e_\Theta(B)$:\\
Since $C \PAR D \not\in e_\Theta(A)$ (resp. $C \PAR D \not\in e_\Theta(B)$), 
We can easily see that $C \not\in e_\Theta(A)$ 
(resp $C \not\in e_\Theta(B)$), 
since if $C \in e_\Theta(A)$ 
      (resp. $C \in e_\Theta(B)$), 
      then $C$ 
      is a conclusion of $e_\Theta(A)$ 
      (resp. $e_\Theta(B)$). 
      Let $S_{B}$ be a principal switching for $B$ in $\Theta$. 
      Since $C \not\in e_\Theta(B)$, 
      there is the unique path $\theta_1$ from $B$ to $C$ 
      in $\Theta_{S_{B}}$ 
      such that $\theta_1$ passes immediately below $B$.
  Then we have two cases:\\
      (b-1-2-1) \  The case where $\theta_1$ includes $A$:\\
	There is the unique path $\theta_2$ from $A$ from 
        $D$ in $\Theta_{S_{B}}$. 
        Let $\theta'_1$ be the subpath of $\theta_1$ from $A$ to $C$. 
        Let $S'$ be $S_{B}$ except that 
        the $\PAR$-switch for $L_{2 \PAR}$ is deleted and
        the $\PAR$-switch for ${L'}_{1 \PAR}$ selects $A$. 
        Then $\Theta'_{S'}$ is a cycle $\theta'_1, C \TENS D, {(\theta_2)}^r$
        since $\theta'_1$ and $\theta_2$ are preserved when moving to $\Theta'_{S'}$ from $\Theta_{S_{B}}$.\\
      (b-1-2-2) \ The case where $\theta_1$ does not include $A$:\\
	There is the unique path $\theta_2$ from $A$ from 
        $D$ in $\Theta_{S_{B}}$. 
        Let $\theta'_1$ be the subpath of $\theta_1$ from $A \TENS B$ to $C$. 
        Let $S'$ be $S_{B}$ except that 
        the $\PAR$-switch for $L_{2 \PAR}$ is deleted and
        the $\PAR$-switch for ${L'}_{1 \PAR}$ selects $A$. 
        Then $\Theta'_{S'}$ is a cycle 
        $\theta'_1, C \TENS D, {(\theta_2)}^r, A \TENS B$ 
        since $\theta'_1$ and $\theta_2$ are preserved when moving to $\Theta'_{S'}$ from $\Theta_{S_{B}}$ 
        except $A \TENS B$ is replaced by $A \PAR B$.
\item [(b-1-3)] The case where neither $C$ nor $D$ is a 
            conclusion of $e_\Theta(B)$:\\
      Similar to the case (b-1-1) above.
\item [(b-1-4)] The case where $D$ is neither a conclusion of $e_\Theta(A)$ 
            nor a conclusion of $e_\Theta(B)$:\\
      Similar to the case (b-1-2) above.
\end{enumerate}
\item [(b-2)] The case where $C \PAR D \not\in e_\Theta(A \TENS B)$ 
and $A \TENS B \in  e_\Theta(C \PAR D)$\\
By Proposition~\ref{propEmpireSubsetNeq}, 
$A \TENS B$ is not a conclusion of $e_\Theta(C \PAR D)$ and
$e_\Theta(A \TENS B) \subsetneq e_\Theta(C \PAR D)$.
In this case we easily find a DR-switching $S'$ for $\Theta'$ such that $\Theta'_{S'}$ has a cycle
including $C \TENS D$. 
In the following we prove the claim. 
Let $S_{A \TENS B}$ be a principal switching for $A \TENS B$ in 
$e_\Theta(C \PAR D)$. 
Then we can obtain a principal switching $S_{C \PAR D}$ for $C \PAR D$ 
in $\Theta$ by extending $S_{A \TENS B}$. 
Then the unique path $\theta$ from $C$ to $D$ in $S_{C \PAR D}(\Theta)$ 
includes neither $A, A \TENS B, A$ nor $B,  A \TENS B, A$, 
because in order that $\theta$ includes $A, A \TENS B, A$ or $B,  A \TENS B, A$, 
$\theta$ must enter $e_\Theta(A \TENS B)$ from a conclusion of $e_\Theta(A \TENS B)$ other than $A \TENS B$.
But this is impossible because $S_{C \PAR D}$ is an extension of 
$S_{A \TENS B}$ that is a principal switching for $A \TENS B$. 
Then we have three cases about $\theta$ from $C$ to $D$.
\begin{enumerate}
\item [(b-2-1)] The case where $\theta$ includes neither $A$, $B$, nor $A \TENS B$:\\
Let $S'$ be $S_{C \PAR D}$ except that the $\PAR$-switch for $L_{2 \PAR}$ is deleted and 
the $\PAR$-switch for ${L'}_{1 \PAR}$ selects $A$ or $B$. 
Since $\theta$ from $C$ to $D$ in $S_{C \PAR D}(\Theta)$ survives in $S'(\Theta')$,
$S'(\Theta')$ has a cycle $\theta, C \TENS D, C$. 
\item [(b-2-1)] The case where $\theta$ includes $A,A \TENS B$ or $A \TENS B, A$:\\
Let $S'$ be $S_{C \PAR D}$ except that the $\PAR$-switch for $L_{2 \PAR}$ is deleted and 
the $\PAR$-switch for ${L'}_{1 \PAR}$ selects $A$. 
Since $\theta$ from $C$ to $D$ in $S_{C \PAR D}(\Theta)$ survives in $S'(\Theta')$,
$S'(\Theta')$ has a cycle $\theta, C \TENS D, C$. 
\item [(b-2-2)] The case where $\theta$ includes $B,A \TENS B$ or $A \TENS B, B$:\\
Let $S'$ be $S_{C \PAR D}$ except that the $\PAR$-switch for $L_{2 \PAR}$ is deleted and 
the $\PAR$-switch for ${L'}_{1 \PAR}$ selects $B$. 
Since $\theta$ from $C$ to $D$ in $S_{C \PAR D}(\Theta)$ survives in $S'(\Theta')$,
$S'(\Theta')$ has a cycle $\theta, C \TENS D, C$. $\,$ 
$\Box$
\end{enumerate}
\end{enumerate}
\end{enumerate}
\end{enumerate}
\end{flushleft}

\section{Proof of Lemma~\ref{lemmaMain}}
\label{secLemmaMain}
In this section, we prove Lemma~\ref{lemmaMain}
by proving the following generalized main lemma by induction.

\begin{lemma}[Generalized Main Lemma]
\label{lemmaGeneralizedMain}
Let $\Theta$ be an MLL proof net with a conclusion $C_{0} \PAR D_{0}$ with the $\PAR$-link $L_{\PAR 0}:\frac{C_{0} \quad D_{0}}{C_{0} \PAR D_{0}}$.
We assume that $m_1 \, \, \TENS$-links $L_{\TENS 1}:\frac{A_1 \quad B_1}{A_1 \TENS B_1}, \ldots, L_{\TENS m_1}:\frac{A_{m_1} \quad B_{m_1}}{A_{m_1} \TENS B_{m_1}}$ and
$m_2 \, \, \PAR$-links $L_{\PAR 1}: \frac{C_1 \quad D_1}{C_1 \PAR D_1}, \ldots, L_{\PAR m_2}:\frac{C_{m_2} \quad D_{m_2}}{C_{m_2} \PAR D_{m_2}}$ occur in $\Theta$, where $m_1, m_2 \in \mathbb{N}$. 
Moreover we assume that 
(a) $\Theta_{i,j}=\tpex(\Theta, L_{\TENS i}, L_{\PAR j})$ 
is not an MLL proof net for each $i, j \,  (1 \le i \le m_1, 0 \le j \le m_2)$.
Moreover
we define $\Theta'$ as follows:
\[ \Theta' \equiv_{\mathdef} \tpex(\Theta, \langle L_{\TENS 1}, \ldots, L_{\TENS m_1} \rangle , 
                            \langle L_{\PAR 0}, L_{\PAR 1}, \ldots, L_{\PAR m_2} \rangle ) \]
Then $\Theta'$ is not an MLL proof net. 
\end{lemma}

\begin{figure}[htbp]
\begin{center}
\includegraphics[scale=.8]{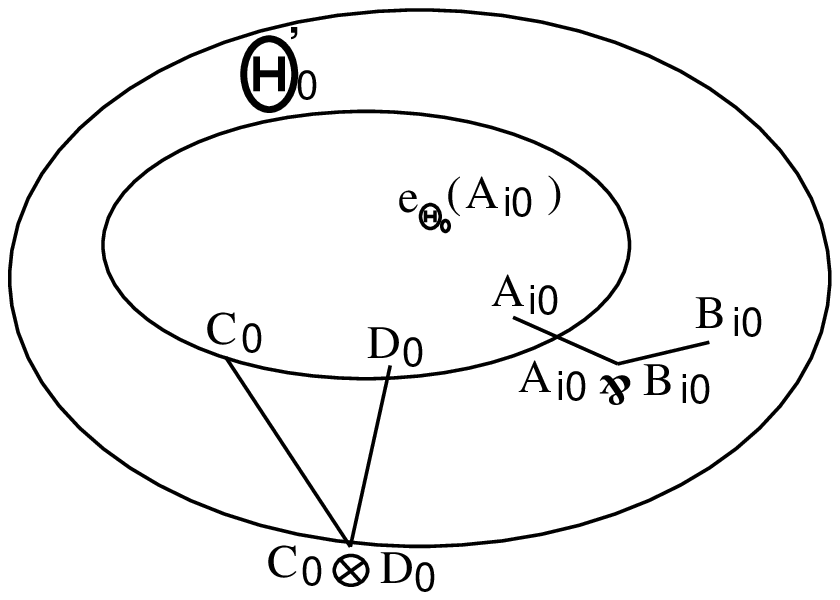}
\caption[$\Theta'$]{$\Theta'$}  
\label{exFigMainLemma}
\end{center}
\end{figure}

\begin{flushleft}
{\it Proof of Lemma~\ref{lemmaGeneralizedMain}} \ 
Let $\Theta_0$ be the MLL proof net obtained from $\Theta$ by deleting $L_{\PAR 0}: \frac{C_{0} \quad D_{0}}{C_{0} \PAR D_{0}}$.
Moreover let $\Theta'_0$ be
\[ \tpex(\Theta_0, \langle L_{\TENS 1}, \ldots, L_{\TENS m_1} \rangle , 
                            \langle L_{\PAR 1}, \ldots, L_{\PAR m_2} \rangle ). \]
We prove the lemma by induction on lexicographic order $\langle m_1, |{\mathbb{L}}_{\Theta}| \rangle$,
$|{\mathbb{L}}_{\Theta}|$ is the number of link occurrences in $\Theta$. 
If $m_1 =0$, then we can easily see that there is a DR-switching $S'_0$ for $\Theta'_0$ such that 
            there is a path $\theta'$ from $C_{0}$ to $D_{0}$ in $S'_0(\Theta'_0)$. 
Therefore $S'_0(\Theta')$ has a cycle. \\
In the following, we prove the induction step: we assume $m_1 > 0$. 
\begin{itemize}
\item The case where $m_2 = 0$:\\
Since $m_1 > 0$, it is obvious that 
there is a DR-switching $S'$ for $\Theta'_0$ such that $S'(\Theta'_0)$ is disconnected.
If $S'(\Theta'_0)$ has more than two maximally connected components, then we have done.
If $S'(\Theta'_0)$ has exactly two maximally components, then $m_1=1$. 
Therefore by condition (a), $\Theta'$ is not an MLL proof net. 
\item The case where $m_2 > 0$:\\
By inductive hypothesis, $\Theta'_0$ is not an MLL proof net. 
Therefore, there is a DR-switching $S'$ for $\Theta'_0$ such that $S'(\Theta'_0)$ has a cycle or is disconnected. 
If $S'(\Theta'_0)$ has a cycle, then we have done: $S'(\Theta')$ also has a cycle. 
If $S'(\Theta'_0)$ has more than two maximally connected components, then we have done: 
$S'(\Theta')$ is disconnected. 
Therefore we can assume that $S'(\Theta'_0)$ has exactly two maximally connected components 
in which each component does not have any cycle 
for any DR-switching $S'$ for $\Theta'_0$ (note that the number of the edges of $S'(\Theta'_0)$ is always the same 
for any DR-switching $S'$ for $\Theta'_0$). 
Hence there is $i_0 \, (1 \le i_0 \le m_1)$ such that 
one connected component has $A_{i_0}$ and the other has $B_{i_0}$. 
Moreover, 
since $\tpex(e_{\Theta_0}(A_{i_0}), \langle L_{\TENS 1}, \ldots, L_{\TENS m_1} \rangle , 
                            \langle L_{\PAR 1}, \ldots, L_{\PAR m_2} \rangle )$
is a subproof structure of $\Theta'_0$, $S'(\tpex(e_{\Theta_0}(A_{i_0}), \langle L_{\TENS 1}, \ldots, L_{\TENS m_1} \rangle , 
                            \langle L_{\PAR 1}, \ldots, L_{\PAR m_2} \rangle ))$ 
must be acyclic and connected for any DR-switching $S'$ for $\Theta'_0$. 
Therefore 
$\tpex(e_{\Theta_0}(A_{i_0}), \langle L_{\TENS 1}, \ldots, L_{\TENS m_1} \rangle , 
                            \langle L_{\PAR 1}, \ldots, L_{\PAR m_2} \rangle )$ is an MLL proof net. 
Therefore by inductive hypothesis, 
$\tpex(e_{\Theta_0}(A_{i_0}), \langle L_{\TENS 1}, \ldots, L_{\TENS m_1} \rangle , 
                            \langle L_{\PAR 1}, \ldots, L_{\PAR m_2} \rangle ) = e_{\Theta_0}(A_{i_0})$
(this means $e_{\Theta_0}(A_{i_0})$ has neither $\TENS$-link nor $\PAR$-link to be exchanged). 
Moreover since by the condition (a) and $S'(\Theta'_0)$ has exactly two maximally connected components for any $S'$, 
for any $j \, (1 \le j \le m_2)$, $e_{\Theta_0}(A_{i_0})$ has neither $C_{j}$ nor $D_{j}$ as a conclusion
(otherwise, the condition (a) is violated, i.e., $e_{\Theta_0}(A_{i_0})$ has $C_{j}$ (resp. $D_{j}$) as a conclusion 
and $e_{\Theta_0}(B_{i_0})$ has $D_{j}$ (resp. $C_{j}$) as a conclusion).
For the same reason,i.e., the condition (a), if $C_{0} ({\rm resp.} \, D_{0}) \in e_{\Theta_0}(A_{i_0})$, then $D_{0} ({\rm resp.} \, C_{0}) \in e_{\Theta_0}(A_{i_0})$ (see Figure~\ref{exFigMainLemma}). 
Then when let $S'_{A_{i_0}}$ be a principal switching for $e_{\Theta_0}(A_{i_0})$ in $\Theta'$, 
$S'_{A_{i_0}}(\Theta')$ is disconnected or has a cycle including $C_{0}, C_{0} \TENS D_{0}, D_{0}$. $\Box$
\end{itemize}
\end{flushleft}

\end{document}